\documentclass{aa}
\usepackage{txfonts}
\usepackage{graphicx}
\usepackage{float}
\usepackage{natbib}
\bibpunct{(}{)}{;}{a}{}{,}

\begin{document}
\titlerunning{IR-Interferometry of the ONC}
\authorrunning{R.Grellmann et al.}
\title{The multiplicity of massive stars in the Orion Nebula Cluster as seen with long-baseline interferometry\thanks{Based on 
observations collected at the European Organisation for Astronomical Research in the Southern Hemisphere, Chile, observing program 086.C-0193}}
\author{R.~Grellmann\inst{1} \and T.~Preibisch\inst{1} \and T.~Ratzka\inst{1} \and S.~Kraus\inst{2} \and K.~G.~Helminiak\inst{3,4} \and H.~Zinnecker\inst{5,6}}
\institute{Universit\"ats-Sternwarte, Ludwig-Maximilians-Universit\"at M\"unchen, Scheinerstr. 1, 81679 M\"unchen, Germany
\and University of Michigan, Department of Astronomy, Ann Arbor, MI 48109-1090, USA
\and Departamento de Astronomía y Astrofisica, Pontificia Universidad Católica de Chile, Av. Vicuña Mackenna 4860, 7820436 Macul, Santiago, Chile
\and Nicolaus Copernicus Astronomical Center, Department of Astrophysics,ul. Rabia\'{n}ska 8, 87-100 Toru\'{n}, Poland
\and SOFIA-USRA, NASA Ames Research Center, Moffett Field, CA 94035, USA
\and Deutsches SOFIA Institut, Universit\"at Stuttgart, 70569 Stuttgart, Germany}
\date{Received / Accepted}

\abstract{The characterization of multiple stellar systems is an important ingredient for testing current star formation models. Stars are more often found in multiple systems,
the more massive they are. A complete knowledge of the multiplicity of high-mass stars over the full range of orbit separations is thus essential to understand their still debated formation process.}
{Infrared long baseline interferometry is very well suited to close the gap between spectroscopic and 
adaptive optics searches. Observations of the Orion Nebula Cluster in general
and the Trapezium Cluster in particular can help to answer the question about the origin and evolution of multiple stars. Earlier studies provide a good knowledge about the 
multiplicity of the stars at very small (spectroscopic companions) and large separations (AO, speckle companions) and thus make the ONC a good target for such a project.}
{We used the near infrared interferometric instrument AMBER at ESOs Very Large Telescope Interferometer to observe a sample of bright stars in the ONC. 
We complement our data set by archival NACO observations of $\theta^1$~Ori~A to obtain more information about the orbit of the close visual companion.}
{Our observations resolve the known multiple systems $\theta^1$~Ori~C and $\theta^1$~Ori~A and provide new orbit points, which confirm the predicted orbit and the determined stellar parameters for $\theta^1$~Ori~C.
Combining AMBER and NACO data for $\theta^1$~Ori~A we were able to follow the (orbital) motion of the companion from 2003 to 2011.
We furthermore find hints for a companion around $\theta^1$~Ori~D, whose existence has been suggested already before, and a previously unknown companion to NU~Ori.
With a probability of $\sim90$\% we can exclude further companions with masses of $\geq3M_\odot$ around our sample stars for separations between $\sim2$\,mas and $\sim110$\,mas.}
{We conclude that the companion around $\theta^1$~Ori~A is most likely physically related to the primary star and not only
a chance projected star. The newly discovered possible companions further increase the multiplicity in the ONC. For our sample of two O and three B-type stars we find on average 
2.5 known companions per primary, which is around five times more than for low-mass stars.}
\keywords{Techniques: interferometric - binaries (including multiple): close - Stars: formation - Stars: individual:$\theta^1$~Ori~C, $\theta^1$~Ori~A, 
$\theta^1$~Ori~D, NU~Ori - Stars: massive}

\maketitle

\section{Introduction}
A high number of stars in our galaxy are found to be in binaries or higher order multiple systems \citep{Mayor1991, Raghavan2010}. The binary frequency is correlated with
the stellar mass and is found to be higher the more massive the star is \citep{Preibisch2001, Delgado2004, Bate2009}. 
Very recently, \citet{Chini2012} found that more than 80\% of the stars with a mass greater than $16\,\rm{M_{\odot}}$ form close binary systems
and \citet{Sana2012} found that over $\sim70$\% of stars born as O-type star will interact with a companion.
Furthermore, the number of companions per system also increases with increasing stellar mass, i.e., massive stars are
often found in triple or higher order systems \citep{Zinnecker07}. Hence, a key step in advancing our understanding of the star and cluster formation process itself 
is an accurate characterization of the properties of binaries as a function of mass and environment. The results of \citet{Kobul2007} and \citet{Chini2012}, e.g., show that the majority of massive stars has close companions with
similar mass, pointing to a multiplicity originating from the formation process.

Recent simulations of star cluster formation by \citet{Bate2009} and \citet{Parker2011} make detailed predictions about the multiplicity of the forming stars and suggest the frequent formation of so-called mini-clusters.
This is, however, observationally still unproven even for the early stages of star formation.   
The observed properties of multiple systems (separation, mass ratio distribution) can be compared with such simulations and thus provide important and strong constraints on star-formation theories. 
For such a comparison, one would ideally like a complete description  of the `primordial' binary population (i.e., a sample of companion stars over the full range of possible orbital distances, that extends from a few stellar 
radii to several 1000~AUs). Another important application of binary observations is the determination of fundamental stellar parameters by following the orbital motion of the system. This is in particular true for the 
determination of stellar masses of young stars, which are needed to calibrate the still uncertain evolution models for the early stages of stellar evolution.

The Orion Nebula Cluster (ONC), a part of the Orion OB1 association, located at a distance of 414~pc \citep{Menten2007}, is a very good target for such a multiplicity study. It contains $\sim3500$ young ($\leq 10^6$\,yrs) stars, 
of which $\sim30$ are O- and B-stars \citep{Hillenbrand1997}. 
While many lower mass stars are still pre-main sequence (PMS) stars, the high-mass stars of the Trapezium system $\theta^1$~Ori (A, B, C, D) have evolved close to the main sequence. Each of the Trapezium stars itself is again multiple.
Thus, conclusions about whether these systems are gravitationally bound can help to determine if the above mentioned predicted ``mini-clusters'' actually exist. 

The multiplicity of the stars in the Orion cluster is already well characterized as far as either very close spectroscopic ($\leq1$\,AU) or relatively wide visual systems ($\geq50$\,AU) are concerned. Searches for spectroscopic binaries among  
ONC members have been performed by, e.g., \citet{Abt1991,Morrell1991,Herbig2006}. Searches for wide visual binaries have been performed with HST imaging \citep{Padgett1997}, speckle holographic observations \citep{Petr1998}, 
and near-infrared adaptive optic observations \citep{Simon1999,Close2012}.
\citet{Weigelt1999} and \citet{1999Preibisch} performed a bispectrum speckle interferometric survey for multiple systems among the O- and B-type Orion Nebula cluster members 
and found 8 new visual companions. A particularly interesting result of these speckle observations was the discovery of a close (33 mas, $\sim 15$ AU) visual companion to $\theta^1{\rm Ori}$\,C, 
the most massive star in the cluster. \citet{Kraus2007, Kraus2009} followed the orbital motion of this system. They used speckle observations, and several interferometric data taken with IOTA and the 
Astronomical Multi-Beam Combiner (AMBER) at ESOs Very Large Telescope Interferometer (VLTI). This allowed them to determine the orbit of this system and to derive fundamental parameters such as the stellar masses and an orbital parallax.

While these searches for spectroscopic and visual companions have already provided us with important information about the multiplicity of those stars, there is still a serious gap in the range of separations covered, 
extending from separations of a few milliarcseconds (mas) [$\sim 1$\,AU] (too wide for spectroscopic detection) to $\sim 100$\,mas [$\sim 50$\,AU] (too close for speckle and adaptive optics studies). 
Infrared long-baseline interferometry is very efficient in finding companions at angular separations 
between $\sim2$ and $\sim100$\,mas; for stellar distances of a few 100\,pc, it
is ideally suited to fill the observational gap between the very close spectroscopic companions and the wide visual companions. Therefore, an interferometric survey of a stellar sample that has already
been searched for spectroscopic as well as wide visual companions can provide the required observational completeness, at least for sufficiently bright companions with flux ratios $\geq0.1$ ($\sim3\,\rm{M_{\odot}}$, see discussion section).

In this paper we present near-infrared interferometric observations of a sample of bright stars in the Orion Nebula Cluster taken with AMBER at the VLTI. The paper is organized as follows: 
The observations and data reduction are described and summarized in Sect.~\ref{Observations}. In Sect.~\ref{Modeling} the models used for the interpretation
of the obtained data are shown and the results are discussed for all of the observed targets. 

\section{Observations and data reduction\label{Observations}}  
The sample selected for the observations with AMBER consisted of all members of the Orion Nebula Cluster which are bright enough to be observed with AMBER (e.g., K magnitude $\leq 5.5$) and the auxiliary telescopes (ATs). 
This leads to a sample of 9 objects. For each target at least two observations are necessary for a reliable determination of the system parameters. 
AMBER \citep{Petrov} is the near-infrared interferometric instrument located at the VLTI and can combine light from  three telescopes in the H- and K-band. For the observations of the ONC targets 
the low resolution mode (LR mode) with a spectral resolution of 35 was used. The ATs have diameters of 1.8~m and can be moved to different stations. For the interferometric observations it is also possible to use the Unit Telescopes
(UTs), which have diameters of 8.2~m, but are on fixed positions. 

The program was granted time in the course of the ESO program 086.C-0193 (P.I.: R.~Grellmann). We also included in our analysis the previous observations of $\theta^1{\rm Ori}$\,C~and~D from the observing run
078.C-0360 (P.I.: S.~Kraus). Further observations were taken as backup targets also in the course of the programs
386.C-0721 (P.I.: R.~Grellmann), 386.C-0650 (P.I.: R.~Grellmann), and 088.D-0241 (P.I.: K.~Helminiak). Therefore, the data were taken with a variety of telescope configurations, which can be found together with information about baselines, position angles, and
calibrators used in Table~\ref{Table_AMBER_Obs}. All data were taken in the LR mode. In total, 6 stars (2 stars of spectral type O, 3 of spectral type B, and one A star) of the 9 sample stars were observed, two of them only once. 
However, some of the observations are of very low quality due to bad weather conditions. These data are not taken into account for the further analysis and discussion. 
As AMBER is a single-mode instrument using optical fibres the field of view is limited to the Airy disk of the individual apertures. This results in a FOV
of 250\,mas for the ATs and 60\,mas for the UTs.

The amdlib software Vers. 3.0.3 \citep{Tatulli, Chelli} provided by the Jean-Marie-Mariotti-Center\footnote{JMMC, http://www.jmmc.fr} 
was used for the data reduction and calibration. As for the detection of a binary signal 
(i.e., a sinusoidal variation in the visibility)
an absolute calibration is not necessary, the data were only calibrated using the associated calibrator rather than a set of calibrators observed over the whole night.
This is a simplified method, assuming that the (instrumental and atmospheric) transfer function has been constant between the star and the calibrator. 
The software accounts for the diameters of the calibrator stars automatically and usually uses the provided database to get the angular sizes.
This was the case for all of the chosen calibrators. Otherwise, they can be specified by the user.

The most critical point for the calibration process is the accurate 
calibration of the wavelength, which is not guaranteed within the amdlib software. 
Thus, we compared the telluric absorption lines in the spectra of the calibrator stars (as they are brighter) with the telluric gaps in 
spectra observed at the Gemini Observatory, similar to the procedure described in \citet{Kraus2007} (but without dividing the object spectrum by the P2vm spectrum) 
and shifted them accordingly where necessary. However, the
uncertainty in wavelength calibration is still a large error source. The maximum achievable accuracy is $0.03\,\mu$m (i.e., one spectral channel), which
for scales to an uncertainty of $\sim 2\%$ \citep{Kraus2009}.

\begin{figure*}
\includegraphics[width=9.0cm, angle=270]{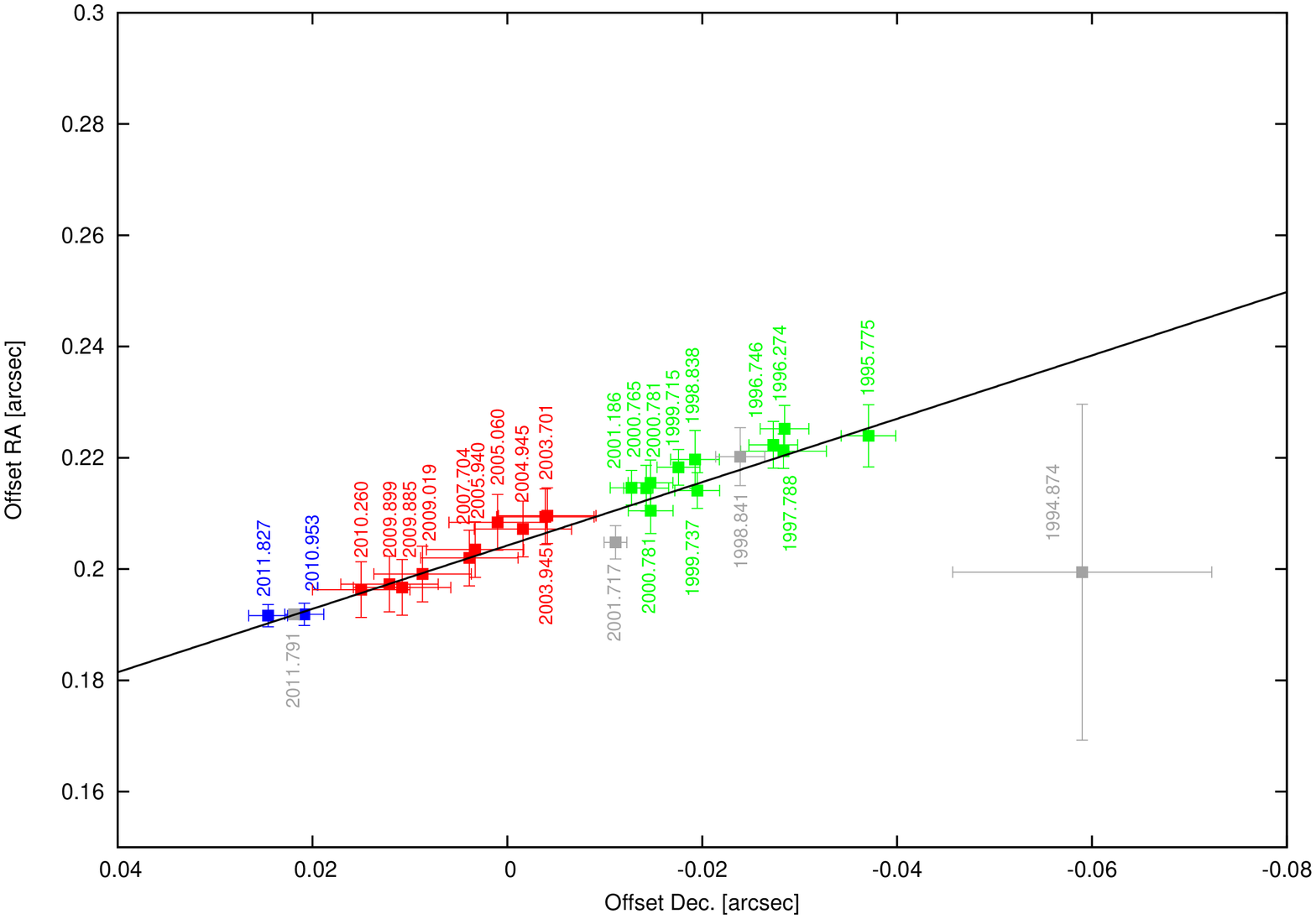}
\caption{\label{Orbit_PAR1865} Positions of the companion of $\theta^1$ Ori A. The blue positions are the new positions 
measured with AMBER, red are the new positions measured with NACO, green are the positions from \citet{Schertl2003},
and the gray points are various other measurements (see Table~\ref{Table_Positions_PAR1865}).}
\end{figure*}

For the supplementing NACO data we searched the ESO archive, where we found a
large number of observations in various filters. The data were reduced with
the instrument-specific ESO pipeline and we selected those images in which
$\theta^1 \rm{Ori\,A} $ is not or only marginally saturated. The relative positions
of all stars were derived with {\tt StarFinder} \citep{diolaiti2000}, which
performs PSF-fitting. In most cases $\theta^1 \rm{Ori\,E}$ served as PSF-reference.
A comparison of the measured positions of the stars with the positions
reported in \citet{mccaughrean} allowed us to derive with {\tt
astrom}\footnote{http://starlink.jach.hawaii.edu/} for each image the detector
orientation and plate scale. 
For this calibration the brightest stars and actual
binaries and outliers were excluded. After identifying obvious outliers by their residuals the fits were repeated. The relative position of the companion of
$\theta^1 \rm{Ori\,A}$ in each epoch is listed in Table~\ref{Table_Positions_PAR1865} . Since we have only
analysed single images, the errors are hard to quantify. We thus show a
conservative error of 5\,mas in Fig.~\ref{Orbit_PAR1865}.
 
\begin{table*}
\caption{\label{Table_AMBER_Obs}AMBER Observations of targets in Orion}
  \begin{tabular}{c c c  l  l  l l}
  \hline \hline
  Target               		&  Other       		& Date        . & \multicolumn{3}{c}{Projected Baselines}                                     & Calibrator(s)\\
   {}                  		&  Name        		& {}            & \multicolumn{3}{c}{}                                                        & {}\\ \hline   
   \object{$\theta^1$ Ori C} 	& PAR 1891,HD 37022     & {}   	& K0-I1 	       & A0-I1			& A0-K0             & {}\\
   {}				&	{}		& 05/10/10	& 44\,m; 8$^{\circ}$  	& 110\,m; -89$^{\circ}$	& 123\,m; 70$^{\circ}$&HD 32613, HD 34137 \\
   {}				& {}			& {}	    	& K0-G1			& G1-A0			& A0-K0			 & {}\\
   {}				& {}			& 26/12/10     & 87\,m; 29$^{\circ}$  	& 88\,m; -65$^{\circ}$ 	& 128\,m;   72$^{\circ}$ & HD 32613, HD 33238\\
   {}				& {}			& 27/12/10     & 84\,m; 22$^{\circ}$  	& 90\,m; -67$^{\circ}$ & 122\,m; 70$^{\circ}$ & HD 32613, HD 33238\\\
{}				&	{}		& {}		&{}			&{}			&	{}		& {}\\
   \object{$\theta^1$ Ori A}  	& PAR 1865,HD37020	& {}	      & K0-I1			& G1-I1			&	G1-K0		& {}\\
   {}				&	{}		& 14/12/10	& 44\,m; 174$^{\circ}$ & 38\,m; -158$^{\circ}$& 79\,m; -172$^{\circ}$   & HD 32613, HD 33238\\
   {}                 	        & {}	 		& 29/10/11     & 44\,m; 171$^{\circ}$ 	& 37\,m; -162$^{\circ}$& 79\,m; -176$^{\circ}$  & HD 40605, HD 33238\\
{}				&	{}		& {}		&{}			&{}			&	{}		& {}\\
    \object{$\theta^1$ Ori D} 	& PAR 1889, HD 37023   & {}		& 	K0-I1		& I1-G1			&	G1-K0		& {}\\
    {}				&	{}		& 14/12/10$^*$   & 43\,m; 180$^{\circ}$& 40\,m; -150$^{\circ}$& 81\,m; -166$^{\circ}$  & HD 33238, HD 47667\\
   {}                 	        &  {}          		 & 29/10/11$^*$ & 43\,m; 177$^{\circ}$ & 41\,m; -147$^{\circ}$& 81\,m; -163$^{\circ}$  & HD 33238\\
   {}                 	        &    {}        		 & 30/10/11    & 43\,m; 175$^{\circ}$  & 38\,m; -156$^{\circ}$& 79\,m; -171$^{\circ}$  & HD 32613\\
   {}				&	{}		& {}		& 	U1-U3		& 	U3-U4		&	U1-U4		& {}\\
    {}                 	        & {}		     	& 17/01/11     & 102\,m; 38$^{\circ}$ 	& 58\,m; -113$^{\circ}$& 130\,m; 64$^{\circ}$  & HD 34137, HD 33238\\
   {}                 	        &     {}        	& 09/01/07$^*$   & 101\,m; 40$^{\circ}$& 43\,m; 125$^{\circ}$ & 113\,m;  62$^{\circ}$  & HD 41547\\
{}				&	{}		& {}		&{}			&{}			&	{}		& {}\\
   \object{$\theta^2$ Ori A} 	& PAR 1993, HD 37041   & {}		& 	A1-K0		&  	A1-G1		&   	G1-K0		& {}\\
{}				& {}			&  01/01/12    & 128\,m; -113$^{\circ}$& 74\,m; 110$^{\circ}$ & 90\,m; -147$^{\circ}$  & HD 34137, HD 40605\\
{}				&	{}		& {}		&{}			&{}			&	{}		& {}\\
   \object{NU Ori	}	& PAR 2074, HD 37061    & {}	    &		K0-G1		& 	G1-A0		&	A0-K0		 & {}\\
   {}				&	{}		& 26/03/11    & 90\,m; -144$^{\circ}$ & 68\,m; -52$^{\circ}$ & 111\,m; -106$^{\circ}$  & HD 32613, HD 40605\\
   {}				& {}			& 27/03/11$^*$ & 89\,m ; -144$^{\circ}$& 60\,m; -45$^{\circ}$ & 99\,m; -107$^{\circ}$  & HD 33238, HD 40605\\
   {}				&	{}		& {}		& 	A1-G1		& 	G1-K0		&	A1-K0		& {}\\
   {}				& {}     		& 31/12/11     & 80\,m; 107$^{\circ}$ & 86\,m; -153$^{\circ}$ & 127\,m; -115$^{\circ}$ & HD 32613, HD 40605\\
{}				&	{}		& {}		&{}			&{}			&	{}		& {}\\
\object{V*~T Ori	}	& PAR 2247, BD-05 1329 & {}		  &	U1-U3		&	U3-U4		 &	U1-U4		 & {}\\
{}				& {}			& 18/01/11    & 102\,m; 40$^{\circ}$ 	& 53\,m; 116$^{\circ}$ & 126\,m; 64$^{\circ}$  & HD 34137\\
{}				& {}			& {}		&{}			&{}			&	{}		& {}\\
\end{tabular}
\tablebib{*Data of low quality and not further mentioned in discussion}
\end{table*}

\section{Modeling\label{Modeling}}
As mentioned above, a typical sign for the existence of a companion (within a certain separation) is a sinusoidal variation of the visibility. 
Describing a binary as composition of two point-sources separated by the distance $a$
the visibility is given by
\[ V(B_{\lambda})= \sqrt{\frac{1+f^2+2f\cos{2\pi aB_{\lambda}}}{{(1+f)}^2}}, \]
with $f$ being the flux ratio of the two sources ($f=I_2/I_1$, $0<f<1$), and $a$ being the distance of the sources after projection on a plane parallel to the baseline $\vec{B}$. The distance of the two sources can then be determined
from the oscillation frequency, whereas the oscillation amplitude depends on their flux ratio. When deriving the separation of the binary, i.e., the position of the companion, using only the visibilities an ambiguity of 
$180^{\circ}$ remains. For a unique solution one needs to consider the closure phase, which is not affected by phase errors due to the earth 
atmosphere and provides information about the (a)symmetry of an object. The closure phase information is thus taken into account to solve the $180^{\circ}$ ambiguity when
determining the positions of the companions of $\theta^1$~Ori~C, $\theta^1$~Ori~A, and $\theta^1$~Ori~D.
The assumption of the sources being point-like is valid as the radius for a star of spectral type B1 is $\sim8\,\rm{R_{\odot}}$. 
For the Orion Nebula Cluster this corresponds to an angular size of $\sim0.1$\,mas. Thus, the stars themselves are unresolved with AMBER.

For the determination of the positions of the companions we used the LitPRO modeling tool developed at the Jean-Marie Mariotti Center \citep{TallonBosc}. 
LitPRO is especially designed to perform model fitting of optical interferometric data. Different geometrical models can be chosen and combined, such as point-sources, 
disks, rings, and different limb-darkening functions.
The fit of the model parameters can be visualized in a $\chi^2$ plane for any two parameters, which enables an easy finding of local and global minima. 

For the determination of the companions positions we used a model of two point-sources and used the separation in RA and DEC as well as the flux ratio
as fitting parameter. Due to problems with the calibrations (i.e., no absolute calibration of the data) the flux ratios determined have high uncertainties and are not
reliable. For the estimation of the errors in separation and PA for $\theta^1$~Ori~C and $\theta^1$~Ori~A we considered the uncertainty in wavelength calibration 
as well as the standard deviations from the LITpro fit.

\subsection{$\theta^1$ Ori C}
The most massive and brightest star in the Orion Nebula Cluster is $\theta^1$~Ori~C. According to its stellar temperature of $T_{\rm eff}=39\,000$\,K \citep{Simon2006} it is of spectral type O7-O5.5. 
$\theta^1$~Ori~C was discovered to be a close visual companion by \citet{Weigelt1999} using bispectrum speckle interferometry. \citet{Kraus2007, Kraus2009} monitored the system to follow its orbital motion
using infrared and visual bispectrum speckle interferometry as well as infrared long-baseline interferometry. They find a total system mass of $44\pm7\,M_{\odot}$ and an orbital period of 11~yrs. 

\begin{figure*}
\includegraphics[width=12.0cm, angle=0]{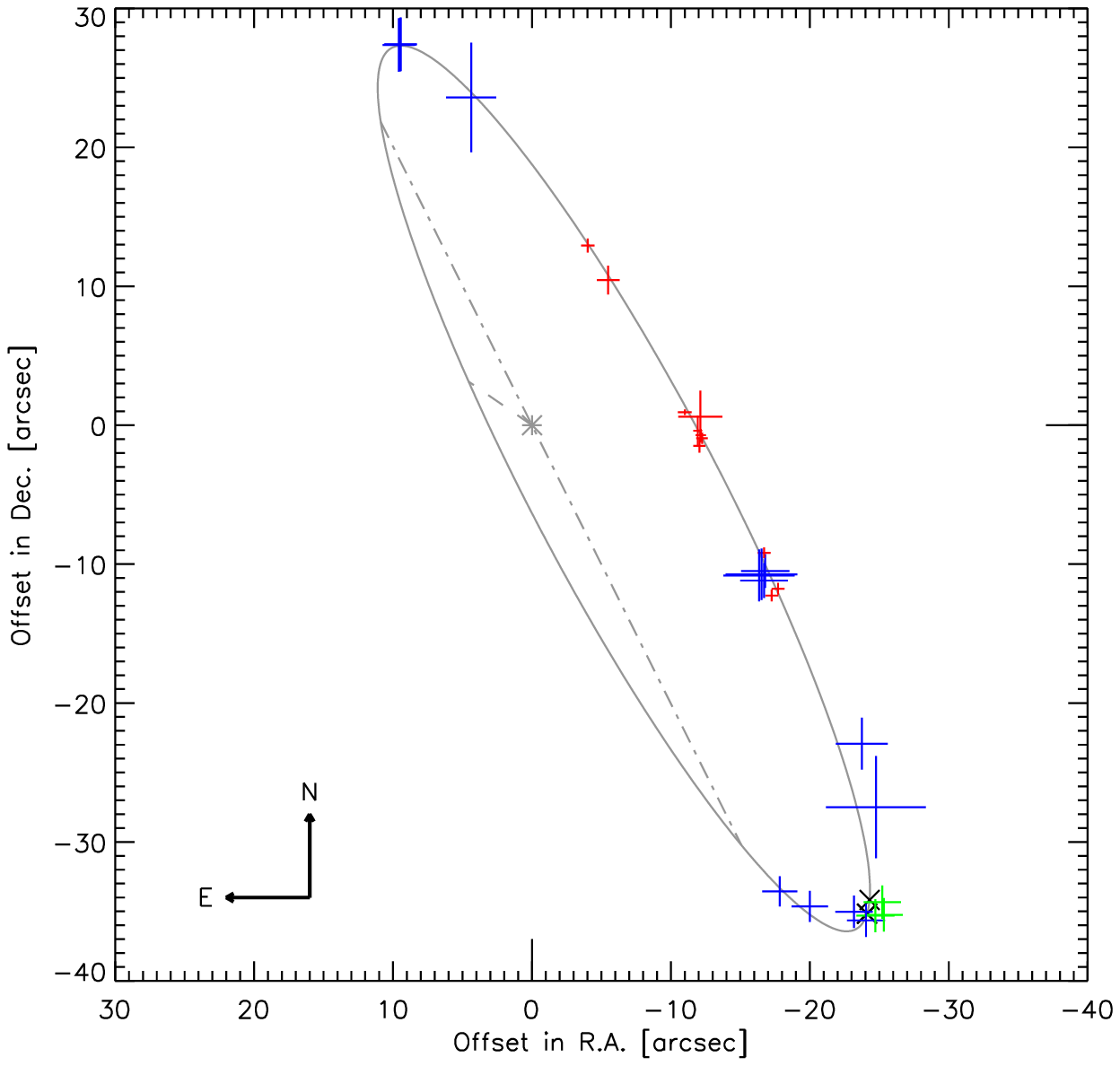}
\caption{\label{Orbit_PAR1891} Orbit of $\theta^1$ Ori C from \citet{Kraus2009}. The green points are the new orbit positions measured with AMBER.}
\end{figure*}

As mentioned above, a good model for binary sources in Orion observed with AMBER is a combination of two point-sources. Fitting the visibilities for the three observations of $\theta^1$~Ori~C, we find a minimum in the 
x-y-coordinate plane at $\sim-25$\,mas in right ascension and $\sim-35$\,mas in declination for all three measurements. The exact positions as well as older orbit points of the close visual companion can be found in
Table~\ref{Table_Positions_PAR1891}. Plotting our positions together with the orbit points used by \citet{Kraus2009} (see Fig.~\ref{Orbit_PAR1891}) we find that the new obtained positions agree very well with the prediction.
This confirms the orbital solution found by \citet{Kraus2009}.

In Table~\ref{AMBER_Targets} it can be seen that there is another very close binary, which due to its angular separation of $1-2.5\,$mas 
\citep{Lehmann2010, Vitrichenko2012} could be resolved with AMBER. However, due to the small mass of this companion (and thus the low flux ratio)
we would need an optimized observing
strategy to be able to detect it in the interferometric data (e.g., assuming a mass of $36\,M_{\odot}$ for the primary and  $1\,M_{\odot}$ for the 
companion we expect the amplitude of the wobbling of the primary to be only 0.03\,mas.). 

\begin{table*}
\caption{\label{Table_Positions_PAR1891}Positions of the close visual companion of $\theta^1$~Ori~C}
 \begin{tabular}{c c c c c c c c}
 \hline \hline
Date & PA & $\Delta$ PA & Diff. Pred. PA* &Sep. & $\Delta$ Sep.& Diff. Pred. Sep.* & Ref. \\
 {} & [${}^{\circ}$] &[${}^{\circ}$] &[${}^{\circ}$] &[$''$] &[$''$] &[$''$] &  {}\\
\hline
1997.784 & 226.0  & 3 & {} & 33 & 2 & {} &1\\
1998.383& 222.0 & 5 & {} & 37 & 4 & {} &1\\
1999.737 & 214.0  & 2 & {} & 43 & 1& {} &2\\
1999.819& 213.5& 2 & {} & 42 & 1 & {}& 3\\
2000.873& 210.0& 2 & {} & 40 & 1 & {}& 3\\
2001.184&  208.0& 2 & {}& 38 & 1 & {} & 2\\
2003.8& 19.3 & 2 & {} & 29 & 2 &{} &  3\\
2003.925& 19.0 & 2 & {} & 29 & 2 & {} & 3 \\
2003.928&19.1 & 2 & {} & 29 &2 & {}& 3 \\
2004.822& 10.5 & 4 & {}& 24 & 4 & {}& 3\\
2005.921& 342.74 & 2 & {}& 13.55 & 0.5 & {} &3\\
2006.149& 332.3 & 3.5 & {}& 11.80 & 1.11 & {} & 4\\
2007.019& 274.9 & 1 & {}& 11.04 & 0.5 & {}& 5\\
2007.143& 268.1& 5.2 & {}& 11.94 & 0.31 & {}& 4\\
2007.151& 272.9 & 8.8 & {}& 12.13 & 1.58 & {}& 4\\
2007.175& 266.6& 2.1 & {}& 12.17 & 0.37 & {}& 4\\
2007.206& 265.6 & 1.9 & {}& 12.28 & 0.41 &{}& 4\\
2007.214& 263.0& 2.3 & {}& 12.14 & 0.43 & {}& 4\\
2007.901& 238.0 & 2 & {}& 19.8 & 2 & {}& 5\\
2007.923& 241.2 & 1 & {}& 19.07 & 0.5 & {}& 5\\
2008.027& 237.0 & 3 & {}&  19.7 & 3 & {}& 5\\
2008.027& 236.5& 3 & {}& 19.6 & 3 & {}& 5\\
2008.071& 236.2 & 2 & {}& 20.1 & 2 & {}& 5\\
2008.148& 234.6& 1 & {}&  21.17 & 0.5 & {}& 5\\
2008.173& 236.4 & 1 & {}& 21.27 & 0.5 &{}& 5\\
2010.762& 216.3 & 2 & +0.9 & 42.6 & 1 & +0.7 &this paper\\
2010.986& 215.7 & 2 & +1.3 & 43.4 & 1 & +0.8  &this paper\\
2010.989& 215.0 & 2 & +0.6 &43.1 & 1 & +0.5  & this paper\\
\end{tabular}
\tablebib{*Differences between observed new orbit positions and positions predicted by the orbit obtained by \citet{Kraus2009},
(1)~\citet{Weigelt1999}; (2)~\citet{Schertl2003}; (3)~\citet{Kraus2007}; (4)~\citet{Patience2008}; (5)~\citet{Kraus2009}}
\end{table*}

\subsection{$\theta^1$ Ori A}
The B0.5 star $\theta^1$~Ori~A was discovered to have a spectroscopic companion at a separation of at $\sim1$\,AU by \citet{Bossi1989} and a close visual companion at a separation of $\sim500$\,AU by \citet{Petr1998}.
\citet{Schertl2003} followed the system's (orbital) motion over several years and found that the relative motion of the companion is consistent with an inclined circular or elliptical orbit, but also 
with a straight line (i.e., a physically unbound, chance projected system). Using the flux ratio from the speckle observations and photometric data compiled by \citet{Hillenbrand1998}, \citet{Schertl2003} estimated the near-infrared
magnitudes, the extinction, and finally the masses. For $\theta^1$~Ori~A1 they found $A_V \approx 1.89$\,mag and $M \approx 16\,\rm{M_{\odot}}$, for $\theta^1$~Ori~A2 they found $A_V\approx3.8$\,mag and $M\approx4\,\rm{M_{\odot}}$. 

We determined new orbit points for $\theta^1$~Ori~A2 using archival NACO observations (see Table~\ref{NACO_PAR1865}) and new AMBER observations. All new positions as well as the positions from older publications can be found in
Table~\ref{Table_Positions_PAR1865}. The position determined from the AMBER data agrees very well with very recent observations taken with the Large Binocular Telescope (LBT) by \citet{Close2012}. Fitting the observations 
with a linear movement, we obtain a relative velocity of $\sim8.5\pm 1$\,km/s. This is in between estimates of $\sim7.2\pm 0.8$\,km/s by \citet{Close2012} and $\sim10.3$\,km/s by \citet{Schertl2003}. 
\citet{Menten2007} observed $\theta^1$~Ori~A2 using the Very Long Baseline Array. They find a proper motion of $9.5$\,km/s in RA and $-3$~km/s in Dec, what leads to a total velocity of $\sim10$\,km/s well
consistent with our velocity measurement. The velocity dispersion of the stars in the ONC has been estimated to be $\approx2.3$\,km/s \citep{Jones1988} assuming a distance of 470\,pc. 
For a distance of 414\,pc this scales to a velocity dispersion of $2.0$\,km/s.
Since the relative velocity of A2 with respect to A1 is more than four times larger than this value it is highly unlikely that we see a random chance projection of unrelated stars.
Furthermore, as discussed in \citet{Schertl2003}, the probability to see a chance projected star with a K-band magnitude of $\leq 9$ 
at an angular separation of $\leq1''$ to the position of A1 is only 0.4\%. Thus, although the observed motion is still linear and shows no significant curvature,
we conclude that it is probably part of an binary orbit seen under a relatively high inclination.

\begin{table}
\caption{\label{NACO_PAR1865}NACO observations of the close visual companion of $\theta^1$~Ori~A}
 \begin{tabular}{c c c c}
 \hline \hline
Date & Obs.-ID & Camera & Band  \\
\hline
2003.7014&060.A-9026(A)& L27& L'\\
2003.9452&072.C-0492(A)& L27&L'\\
2004.9452&074.C-0637(A)&S13&Ks\\
2005.0603&074.C-0401(A)&L27&4.05\\
2005.9397&076.C-0057(A)&S27&2.17\\
2007.7041&079.C-0216(A)&S27&2.17\\
2009.0192&482.L-0802(A)&S27&2.12\\
2009.8849&060.A-9800(J)&S27&Ks\\
2009.8986&084.C-0396(A)&L27&L'\\
2010.2603&085.C-0277(A)&L27&L'\\
\end{tabular}
\end{table}

\begin{table}
\caption{\label{Table_Positions_PAR1865}Positions of the close visual companion of $\theta^1$~Ori~A}
 \begin{tabular}{c c c c c c}
 \hline \hline
Date & PA & $\Delta$ PA &Sep. & $\Delta$ Sep. & Ref. \\
 {} & [${}^{\circ}$] &[${}^{\circ}$] &[$''$] &[$''$] & {}\\
\hline
1994.901 & 343.5 & 5 & 208 & 30 &3\\
1995.775 & 350.6 & 2 & 227& 5 &2\\
1996.247 & 352.8& 2 & 227&4 &2\\
1996.746 & 352.7 & 2& 223& 4&2\\
1997.788 & 353.0 & 2&224 & 4&2\\
1998.838 & 353.8 & 2&221 & 5&2\\
1998.841 & 353.8 & 2 & 221.5 & 5&4\\
1999.715 & 355.4 & 2&219 & 3&2\\
1999.737 & 354.8 & 2&215& 3&2\\
1999.819 & 175.1*& 0.5& 212&2.5&5\\
2000.765 & 356.2 & 2&215& 4&2\\
2000.781 & 356.1 & 2& 216&4 &2\\
2000.781 & 356.0 & 2& 211& 4&2\\
2001.186 & 356.0 & 2& 215& 3&2\\
2001.718 & 356.9 &1 & 205.1 & 3 &1\\
2003.701 & 3.9& 1&210&5&this paper\\
2003.945 & 3.9& 1&209&5&this paper\\
2004.816 & 0.3& 1.6& 203&2& 6\\
2004.822 & 0.9&0.8&205&3&6\\
2004.945 & 4.6& 1&207&5&this paper\\
2005.060 & 5.3& 1&208&5&this paper\\
2005.940 & 5.9& 1&204&5&this paper\\
2007.704 & 6.1& 1&202&5&this paper\\
2009.019 & 7.5& 1&199&5&this paper\\
2009.885 & 8.2& 1&197&5&this paper\\
2009.899 & 8.5& 1&198&5&this paper\\
2010.260 & 9.4& 1&197&5&this paper\\
2010.877 & 6.5 & 0.3&193.1&0.5&1\\
2010.953 & 6.2 & 2& 193.0&1& this paper\\
2011.827 & 7.3 & 2& 193.2&1& this paper\\
\end{tabular}
\tablebib{*$180^{\circ}$ ambiguity could not be solved from the measurement itself, but taking into account the other measurements should probably be $355.1^{\circ}$,
(1)~\citet{Close2012}; (2)~\citet{Schertl2003}; (3)~\citet{Petr1998}; (4)~\citet{Weigelt1999}; (5)~\citet{Balega2004}; (6)~\citet{Balega2007}}
\end{table}

\subsection{$\theta^1$ Ori D}
$\theta^1$~Ori~D is of spectral type B0.5 and has been observed by \citet{Kraus2007} with the Infrared Optical Telescope Array (IOTA) in the H-band and 
using the LBT by \citet{Close2012} in the near-IR narrowband ($2.16\,\mu$m and $1.64\,\mu$m).
The  near-infrared adaptive optic observations carried out by \citet{Simon1999} reveal a very wide optical companion with a distance of $\sim1.4''$, which is
also resolved by the LBT observations. It remains unclear, whether this source is really a physical companion
or just, e.g., a background object. A spectroscopic companion with a period of 40~days was identified by \citet{Vitrichenko2002}.
\citet{Kraus2007} find a significant non-zero closure phase signal in the IOTA measurements  suggesting the presence 
of a companion with a separation of 18.4\,mas and a flux ratio of 0.14, although the uv-coverage and SNR was too low for a detailed characterization.

\begin{figure*}
\parbox{18.5cm}{\parbox{9.0cm}{\includegraphics[width=9.0cm, angle=0]{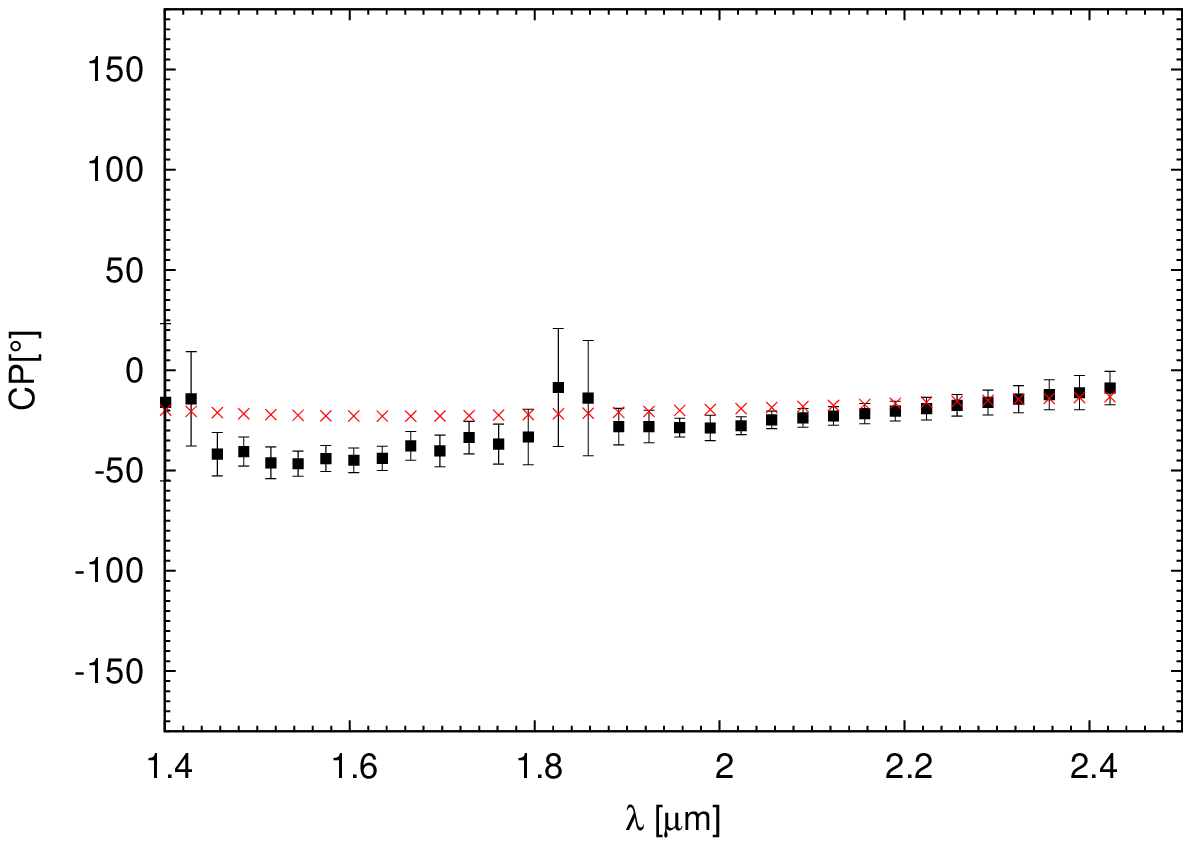}}
\parbox{9.0cm}{\includegraphics[width=9.0cm, angle=0]{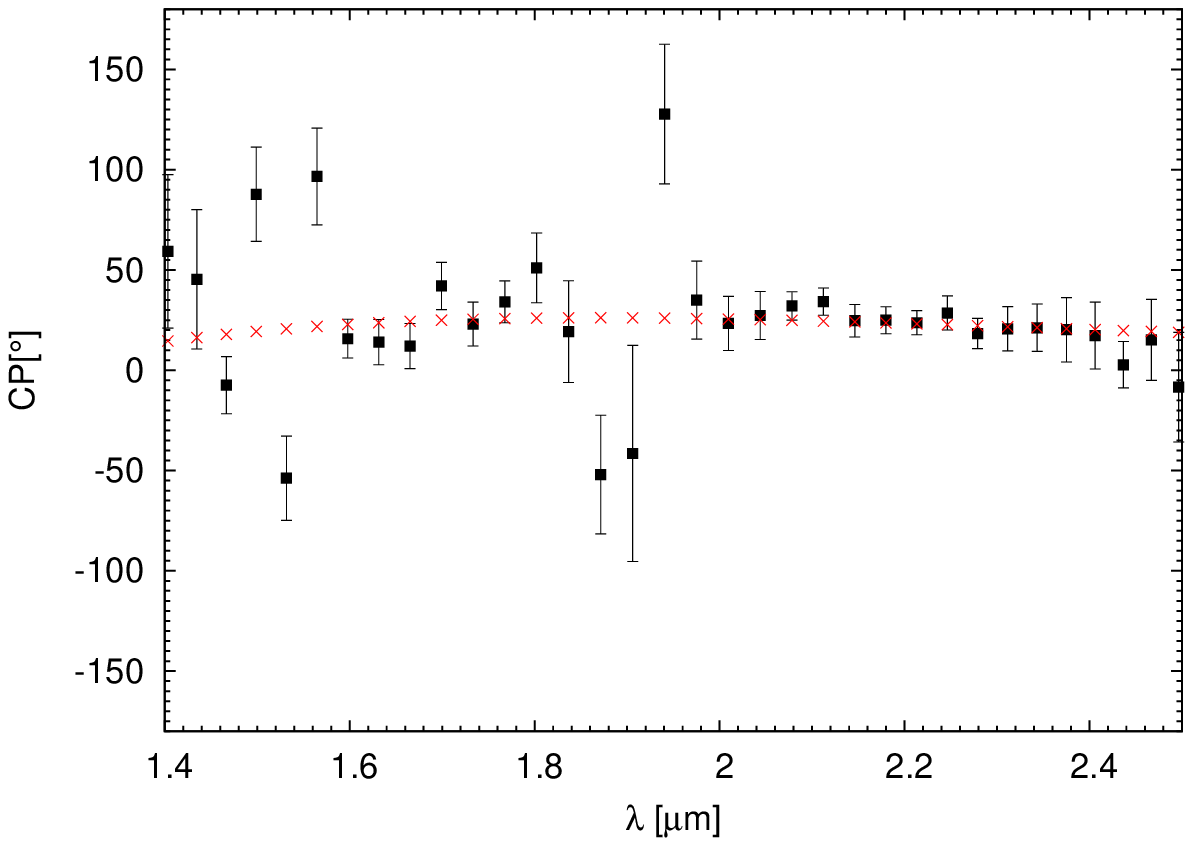}}}
\caption{\label{CP_1889} Closure phases of $\theta^1$~Ori~D from the observations in January 2011 (left) and October 2011 (right) together with the best-fit model as determined with LITpro.}
\end{figure*}

We observed $\theta^1$~Ori~D with AMBER on five different nights 
(see Table~\ref{Table_AMBER_Obs}), however, we can only take into account 2 of the 5 observations (from 30/10/11 and from 17/01/11) as the other ones 
are very noisy. We do not find a clear sinusoidal variation in 
the visibility; however, we a see a clear signal in the closure phase (see Fig.~\ref{CP_1889}), which is consistent with a very close binary. 
In this case, we would expect to see only a slight increase or decrease in the visibility over wavelength band, which is consistent with the observations.
Fitting the data with the LITpro software we find minima for a separation of 
$\sim2\,$mas for the observation on 17/01/2011 and a separation of $\sim4\,$mas
for the observation obtained on 30/10/2011. This component could, considering the large uncertainties in the determination of the separation, be either 
originating from the spectroscopic binary detected by \citet{Vitrichenko2002} or from the additional component 
suggested by \citet{Kraus2007}. However, it seems very unlikely that the spectroscopic binary with a period of only 40\,d is the same
source than the 18.4\,mas binary suggested by the IOTA data.
An alternative explanation for the closure phase signal could be an inclined circumstellar disk,
which would also be an interesting result for a star of spectral type B0.5. 

\subsection{NU Ori}
The B1 star NU Ori is known to have a spectroscopic companion with a period of 8~days \citep{Morrell1991} and a wide visual companion at a distance of $0.47''$ \citep{Koehler2006}.  
Our AMBER observations from 31/12/2011 (see Fig.~\ref{Vis_PAR2074_1}) show a sine-like variation in the K-band, while the H-band data are too noisy to provide useful information. The data
obtained on 26/03/2011 are very noisy on two of the three baselines (see Fig.~\ref{Vis_PAR2074_2}). 
However, there seem to be some oscillations in the visibility data.  For the observations obtained on 31/12/2011 we find a binary model which roughly fits the
visibilities for separation of $\sim20$\,mas, whereas the model for the data from 26/03/2011 favors a separation of $\sim10$\,mas.
The observations thus suggest the presence of a fourth companion in a hierarchical system, but further observations are needed to verify this prediction.

\subsection{Other targets\label{Others}}
\begin{figure*}
\parbox{18.5cm}{\parbox{9.0cm}{\includegraphics[width=9.0cm, angle=0]{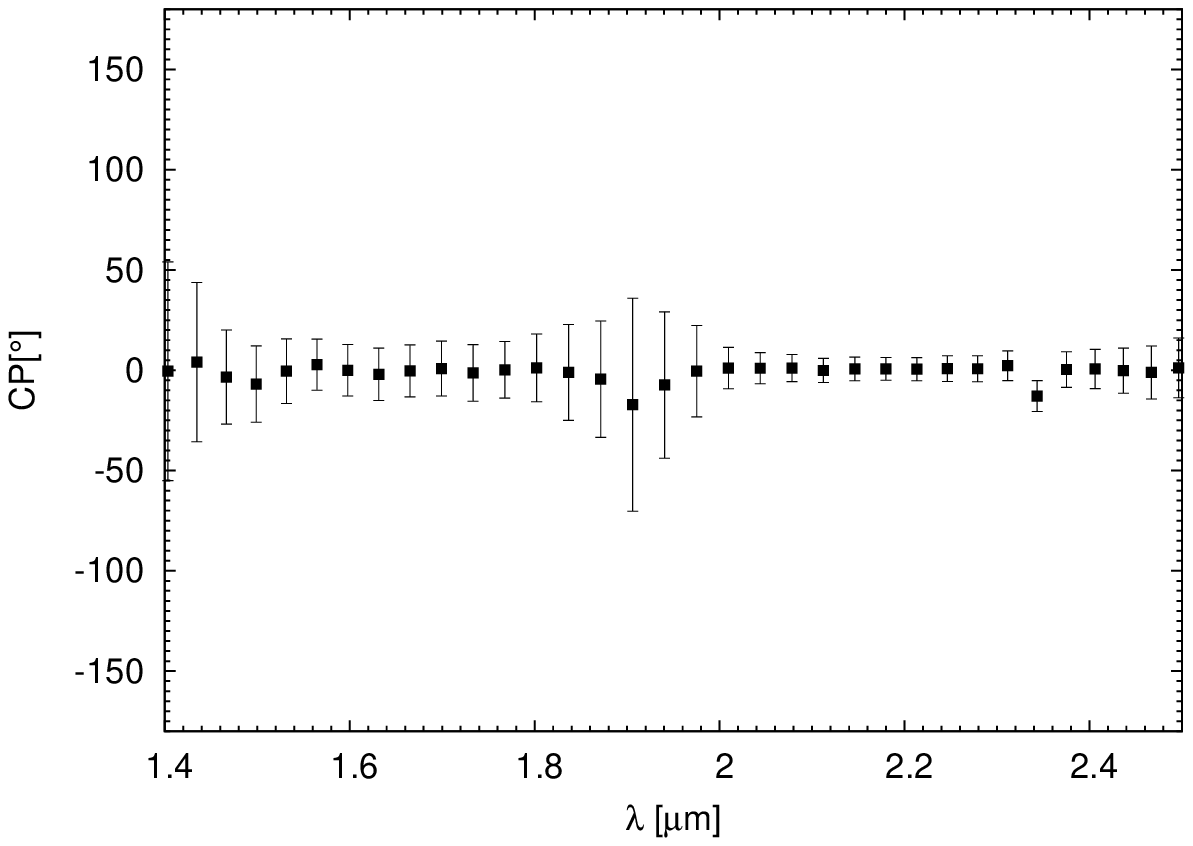}}
\parbox{9.0cm}{\includegraphics[width=9.0cm, angle=0]{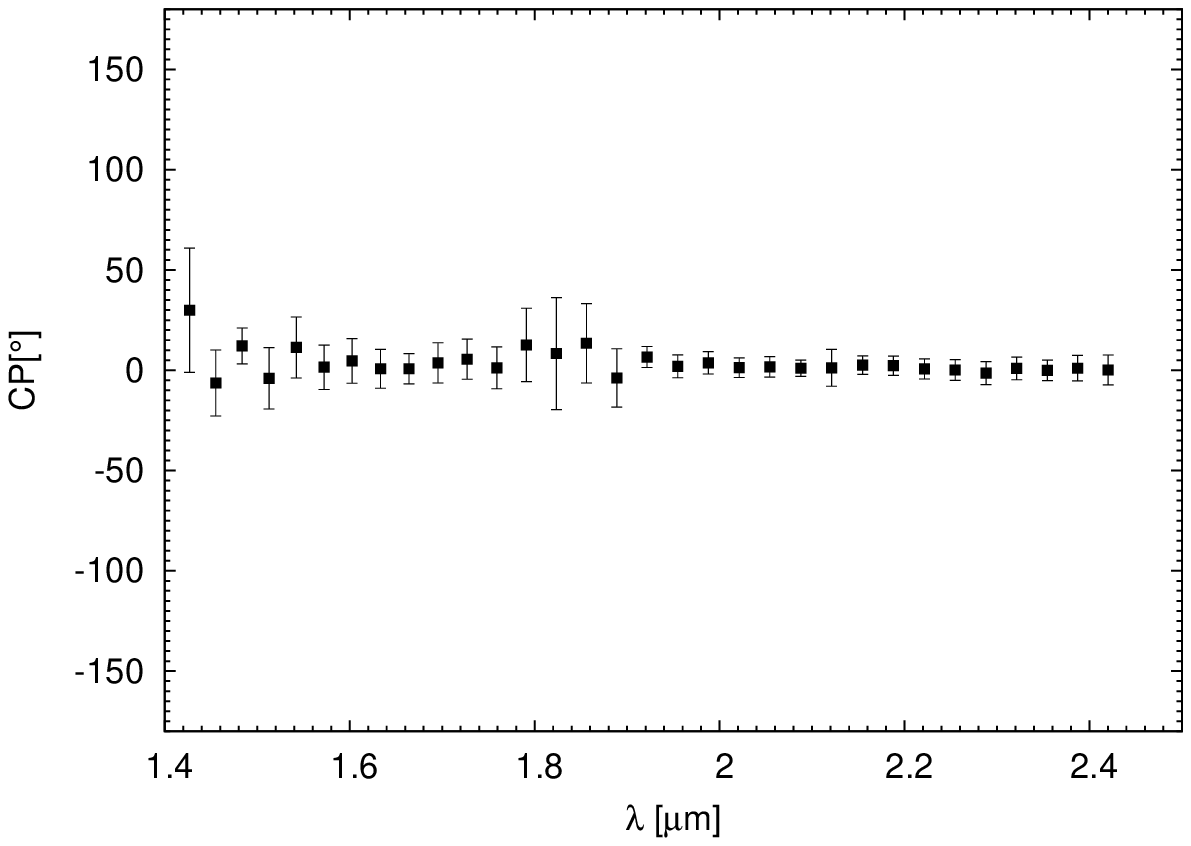}}}
\caption{\label{CP_1993_2247} Closure phases of $\theta^2$~Ori~A and V*~T~Ori.}
\end{figure*}

\begin{figure*}
\parbox{18.5cm}{
\parbox{9.0cm}{\includegraphics[width=9.0cm, angle=0]{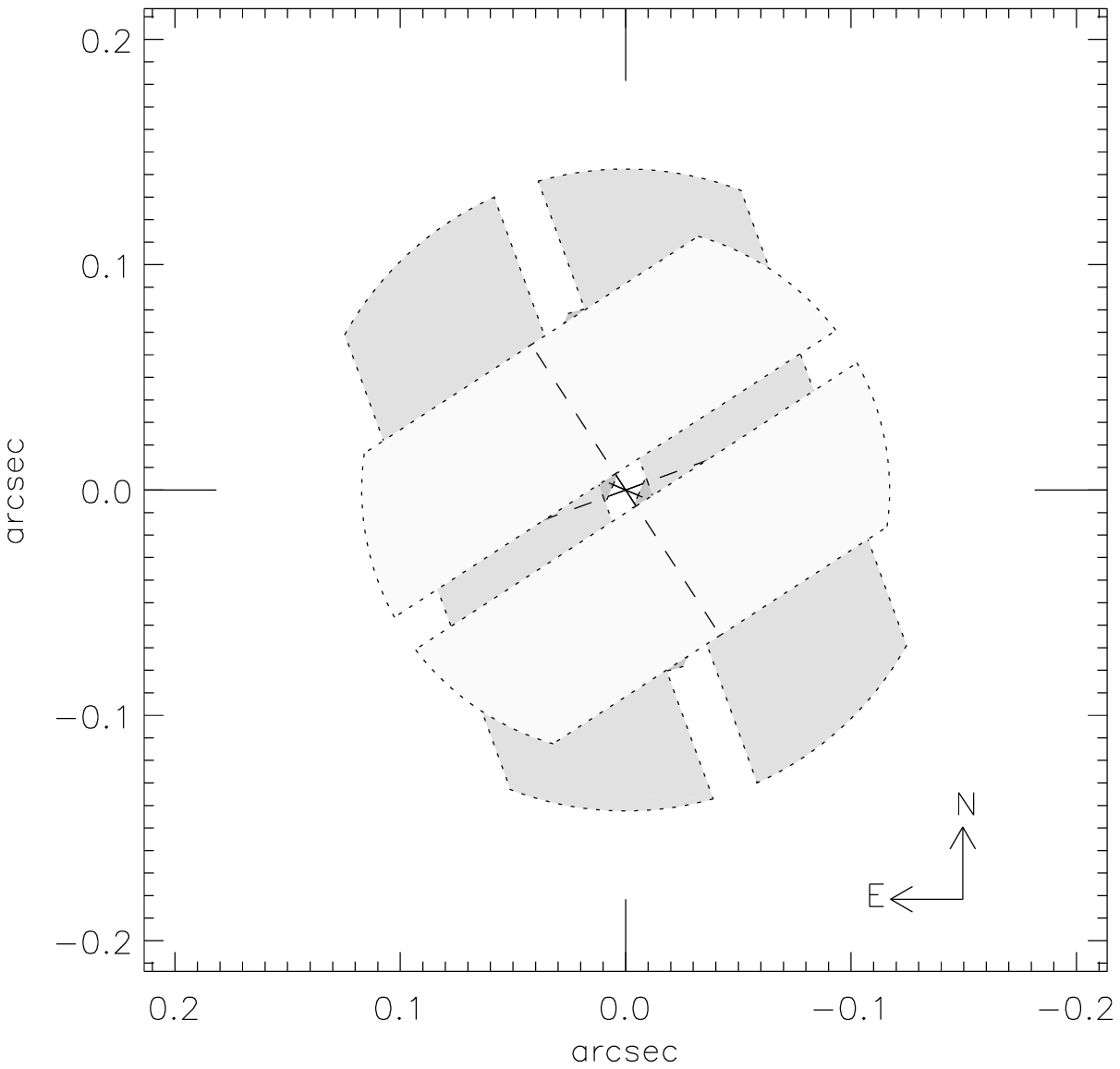}}
\parbox{9.0cm}{\includegraphics[width=9.0cm, angle=0]{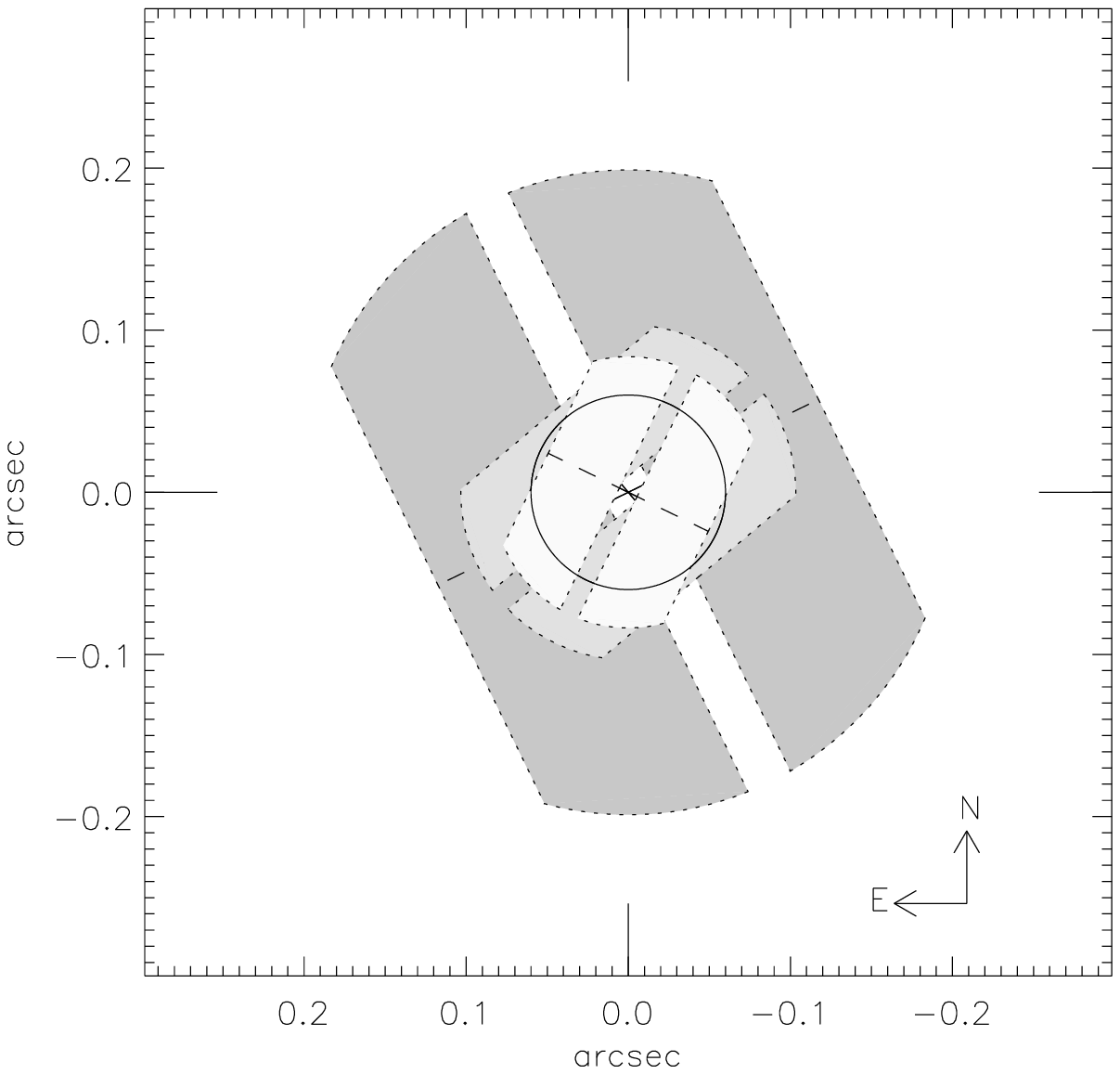}}}
\caption{\label{No_Binaries} 
Coverage of the area around the observed objects, where a binary component within the defined conditions (see Sect.~\ref{Others}) can be excluded from the AMBER data. 
The different colors indicate the areas for the three different baselines and position angles. 
\textit{Left:}  $\theta^2$ Ori A  
\textit{Right:} V* T Ori. Here, the black circle marks the size of the FOV of AMBER with the UTs, thus, a binary component can only be excluded inside this FOV.}
\end{figure*}

In the data for $\theta^2$~Ori~A (spectral type O9.5) and V*~T~Ori (spectral type A3) we do not detect any clear signal for binarity in the visibilities (see Figs.~\ref{Vis_PAR1993_1} and \ref{Vis_PAR2247_1}). The closure phases (see Fig.~\ref{CP_1993_2247}) 
are $\sim0^{\circ}$ and hence do not show any hint for an asymmetry either. 
We can thus define an area in the separation-position angle parameter space, where a sufficiently bright companion can be excluded. For this, we make the following assumption: 
a binary would be detectable in our data if we could see at least half a period of the modulation in one band (i.e., H- or K-band; sometimes we have only flux in the K-band data). 
In the other case (i.e., a very wide binary) at least three data points per modulation are
needed to detect it. Due to the errorbars of the data, we furthermore can only be sure to detect binaries with a flux ratio of $\gtrsim0.1$ (which would result in an amplitude of variation of $\sim0.2$). 
Plots of the areas where a companion can be excluded from the AMBER data can be found in Fig.~\ref{No_Binaries}. It can clearly be seen there that the known $0.38''$ separated companion of $\theta^2$~Ori~A
is too far away already to be detected with AMBER. 
We now can also estimate the probability to miss a companion inside a certain radius, which depends on the baseline configuration (thus it can be different 
for different observations).
The probability to miss a companion around $\theta^2$~Ori~A with a separation between $0.002''$ and $0.11''$ is $\sim10$\%, if it 
is by chance in one of the gaps of the covered area.
To decrease the probability of missing a companion at least one more AMBER 
observation with an appropriate baseline configuration would be needed. 
For V*~T~Ori the separation radius for a detection is limited by the FOV of AMBER together with the UTs. 
We can thus exclude a companion for this source between  $0.002''$ and $0.06''$ (black circle in Fig.~\ref{No_Binaries}). 
 
\begin{table*}
\caption{\label{AMBER_Targets}Companions of the ONC targets. As the observational limits for detectable companions change with the brightness of the primary target and the used baseline configuration, in
row 7 the limits for each of the observed sample stars are given. The mass is the minimum required mass of the companion as obtained from the \citet{Siess2000} models for 
an age of $10^6$\,yrs to be able to be clearly detected. Below the range of separations covered by the observations is given.}
 \begin{tabular}{c c c c c c c c}
 \hline \hline
 Star 			& Component&  Sep./Period& SpT	& Mass 		& Detection Method 	& Comp. detect. limits  &Reference/Comment\\
 {}			& {}	   & [$''$]/days & {}   &[$M_{\sun}$]	& {} 			& AMBER			&{}\\ \hline
  $\theta^1$ Ori C	& C1	& {} 		& O5.5	& $\sim38$ 	& {} 			&   $\sim4\,M_{\sun}$   &2,5\\ 
 {}			& {}	& {}		& {} 	& {}		&{}			& 2--240\,mas 		&{} \\
  {}			& C2	& 0.040$''$	& {} 	&  $\sim5$ 	& speckle interferometry& {}  			&5, this paper\\
  {}			& C3	& 0.002$''$	& {} 	&  $\sim1$ 	& spectroscopy 		&{}			&11,12\\
  {}			& C4	& {}		& {}	&  $\sim1.4$	& spectroscopy		& {}			&10\\
  {}			& C5	& 28 d around C4&{}	&  $\sim1.3$	& spectroscopy		& {}			&10\\
 {}			& {}	& {}		& {} 	& {}		& {}			&{}			&{} \\
  $\theta^1$ Ori A	& A1	& {} 	 	& B0.5 	&  $\sim20$ 	& {} 			& $\sim3\,M_{\sun}$ 	&6\\
 {}			& {}	& {}		& {} 	& {}		&{}			& 3--280\,mas 		&{} \\
  {}			& A2	& 0.2$''$	& {} 	&  $\sim4$ 	& AO 			& {}			&2,3,4, this paper\\
 {}			& A3	& 65 d\,$\approx1.5$\,mas& A0 	&  $\sim4$ & spectroscopy	&{} 			&1,13,14,15,16\\
 {}			& {}	& {}		& {} 	& {}		& 	{} 		&{}			&{}\\
  $\theta^1$ Ori D	& D1	& {}		& B0.5 	& {} 		& {}			& $\sim3\,M_{\sun}$ &	4\\
 {}			& {}	& {}		& {} 	& {}		&{}			& 2--240\,mas 		&{} \\
 {}			& D2	& 1.401$''$	& {} 	& {} 		& AO  			& {}			&4\\
 {}			& D3	& 0.0184$''$	& {}	& {} 		& interferometry	& {}			&4, this paper\\
 {}			& D4	& 40 d		& {}	& {}		& spectroscopy		& {}			& 22, this paper\\
 {}			& {}	& {}		& {} 	& {}		& {} 			&{}			&{}\\
  $\theta^2$ Ori A 	& A1a	& {}		& O9.5 	&  $\sim25$ 	& {} 			& $\sim3.5\,M_{\sun}$ 	&6,7\\
 {}			& {}	& {}		& {} 	& {}		& 			& 2--140\,mas		&{} \\
  {}			& A1b	& 21 d, $\approx1$\,mas	& {} &  $\sim9$& spectroscopic		&{} 			&8,17\\
  {}			& A2	& 0.38$''$	& {}   	&  $\sim7$  	& AO 			&{} 			&7,18\\
 {}			& {}	& {}		& {} 	& {}		& {}			&{} 			&{}\\
  NU Ori		& Aa	& {}		& B1   	& {} 		& {}			& $\sim2.7\,M_{\sun}$ 	&6\\
 {}			& {}	& {}		& {} 	& {}		&{}			& 2--150\,mas 		&{} \\
  {}			& Ab	& 8 -19 d\,$\approx0.4--0.8$\,mas& {}&  $\sim3$& spectroscopic&{} 			&6,8,17\\
  {}			& B	& 0.47$''$	& {}   	&  $\sim1$  	& AO			&{}  			&9\\
  {}			& C	& $\sim$0.015$''$& {}   & {} 		& interferometry	&{} 			&this paper\\
 {}			& {}	& {}		& {}   	& {}		& {}			&{} 			&{}\\
  V* T Ori		& Aa	& {}		& A3   	& {} 		& {} 			&$\sim2.2\,M_{\sun}$	&9\\
  {}			& {}	& {}		& {} 	& {}		&{}			& 2--200\,mas 		&{} \\
  {}			& Ab	& 14.3 d	& {}	&  {}		&spectroscopy		& {}			&20,21\\
  {}			& B	& 0.84$''$	& {}	& {}		& spectroscopy		& {}			&19\\
\end{tabular} 
\tablebib{(1)~\citet{Bossi1989}; (2)~\citet{Schertl2003}; (3)~\citet{Petr1998}; (4)(1)~\citet{Close2012}; (5)~\citet{Kraus2009}; (6)~\citet{Stelzer2005}; (7)~\citet{1999Preibisch}; (8)~\citet{Morrell1991};
(9)~\citet{Koehler2006}; (10)~\citet{Vitrichenko2011}; (11)~\citet{Lehmann2010}; (12)~\citet{Vitrichenko2012}; (13)~\citet{Vitrichenko2001}; (14)~\citet{Vitrichenko1998};
(15)~\citet{Lohsen1975}; (16)~\citet{Lohsen1976}; (17)~\citet{Abt1991}; (18)~\citet{Mason2009}, (19)~\citet{Wheelwright2010}; (20)~\citet{Shevchenko1994};
(21)~\citet{Corporon1999}; (22)~\citet{Vitrichenko2002}}
\end{table*}

\section{Summary \& Conclusions\label{Summary}}
We presented observations of a brightness-selected sample of stars in the Orion Nebula Cluster obtained with the near-infrared interferometric instrument AMBER at the VLTI. We re-observed the already known companions around
$\theta^1$~Ori~C at a distance of $\sim40$\,mas and around $\theta^1$~Ori~A at a separation of $\sim0.2''$. The new orbit points for $\theta^1$~Ori~C
confirm the predicted orbital period of $\approx11$\,yrs and the stellar parameters derived from the fit.
Combining the AMBER data with archival NACO data we can follow the motion of the companion of $\theta^1$~Ori~A. The motion is, however, still consistent with a 
linear movement, no curvature is detected in the trajectory. The relative velocity of $\sim8.5\pm 1$\,km/s  obtained from a linear fit is significantly higher than the velocity dispersion measured for the ONC, such that the observed
motion is probably due to a companion with highly inclined orbit. 
For two of our targets, $\theta^1$~Ori~D and NU~Ori, we find hints for the presence of further companions, which need to be confirmed by further 
observations. 

The detection limits of the AMBER observations depend on the brightness of the primary and the baseline configuration used. In Table~\ref{AMBER_Targets} the covered range in angular separation and the
required minimum mass of the companion are given for each of the observed targets. 
As mentioned in Sect.~\ref{Others} we assume a minimum flux ratio of $\sim0.1$. This corresponds to a magnitude difference of $\Delta\rm{K}\sim2.5$\,mag. 
Using the pre-main sequence models from \citet{Siess2000} for an age of $10^6$\,yrs we can calculate the minimum required companion mass to be detectable.
With the AMBER observations we are able to detect companions with masses down to $\sim3\,M_{\sun}$ and with projected angular
separations of $\sim 1$ to $\sim80\,$AU.

The multiplicity in the Orion Nebula Cluster has been measured and discussed in various publications, e.g., \citet{Hillenbrand1997, 1999Preibisch, Koehler2006}.
\citet{1999Preibisch} find a companion frequency of $\geq1.5$ per primary for the massive stars (earlier than B3) in the ONC, which is around three times higher than for low-mass stars. 
Including all newly discovered possible companions the multiplicity in the ONC increases further. It might be even higher in reality as all of the techniques used to find companions miss, e.g., very faint companions.
For our sample of two O and three B-type stars we find on average 2.5 known companions per primary, around five times more than for low-mass stars. 
This is in agreement with the finding that stars are more often found in multiple systems,
the more massive they are.

\begin{acknowledgement}
We gratefully acknowledge funding of this work by the German \textit{Deutsche Forschungsgemeinschaft, DFG} project number PR~569/8-1.
K.G.H. acknowledges support provided by the Proyecto FONDECYT Postdoctoral No. 3120153 and the Polish Nacional Science Center grant
no. 2011/03/N/ST9/01819.
This research has made use of the Jean-Marie Mariotti Center \texttt{AMBER data reduction package}\footnote{Available at http://www.jmmc.fr/amberdrs} 
and \texttt{LITpro}\footnote{LITpro software available at http://www.jmmc.fr/litpro}
service co-developed by CRAL, LAOG and FIZEAU. This research has made use of the SIMBAD database, operated at CDS, Strasbourg, France. 
We thank the referee Douglas Gies for the helpful und constructive comments, which helped to improve the paper.
\end{acknowledgement}

\bibliographystyle{bibtex/aa}
\bibliography{bibtex/mybib}

\appendix
\section{Visibilities}
\begin{figure*}
\parbox{18.5cm}{\parbox{6.0cm}{\includegraphics[width=6.0cm, angle=0]{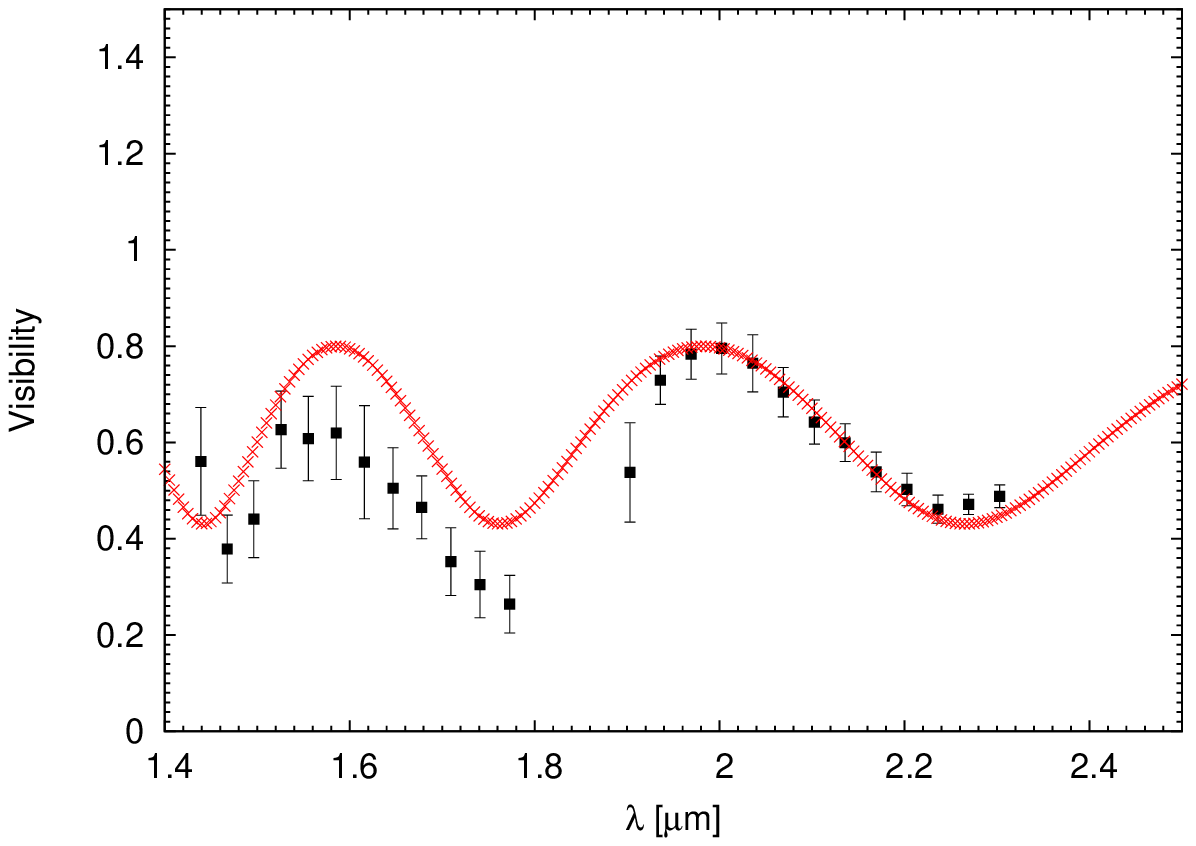}}
\parbox{6.0cm}{\includegraphics[width=6.0cm, angle=0]{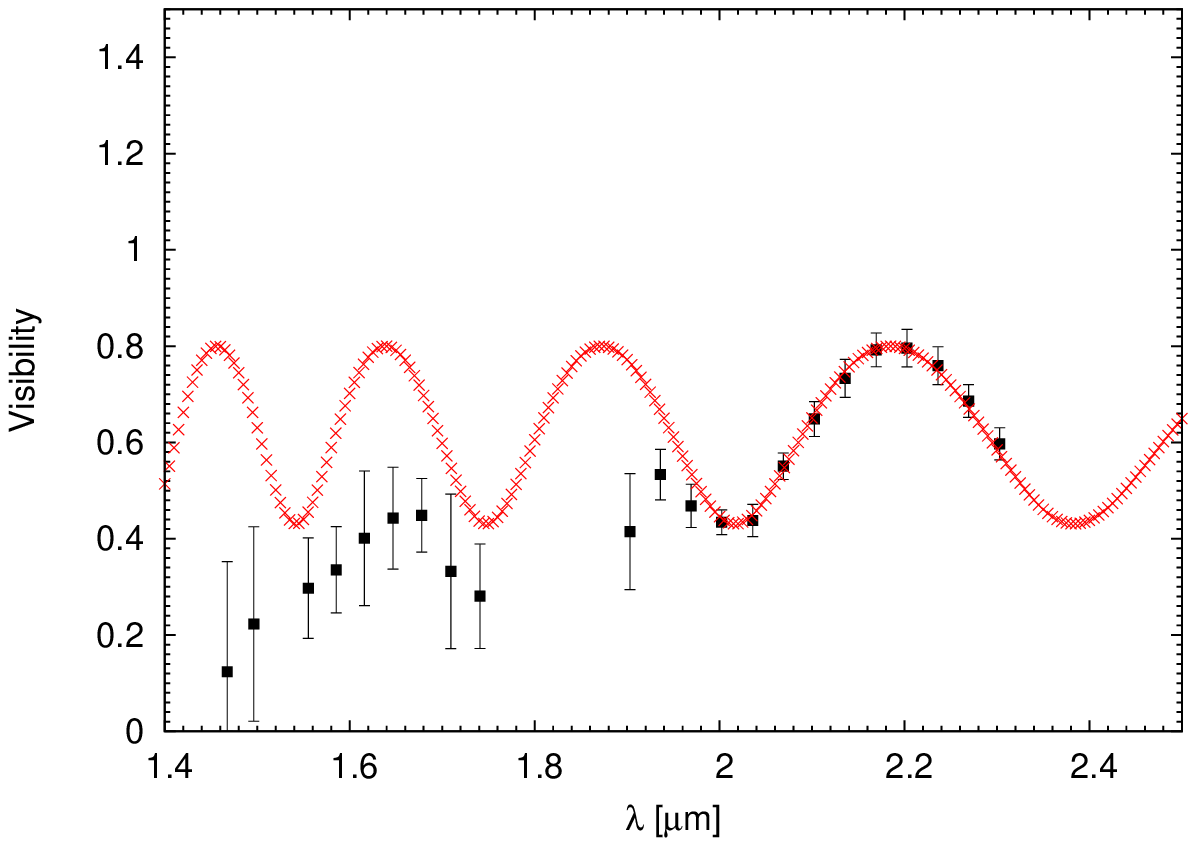}}\parbox{6.0cm}{\includegraphics[width=6.0cm, angle=0]{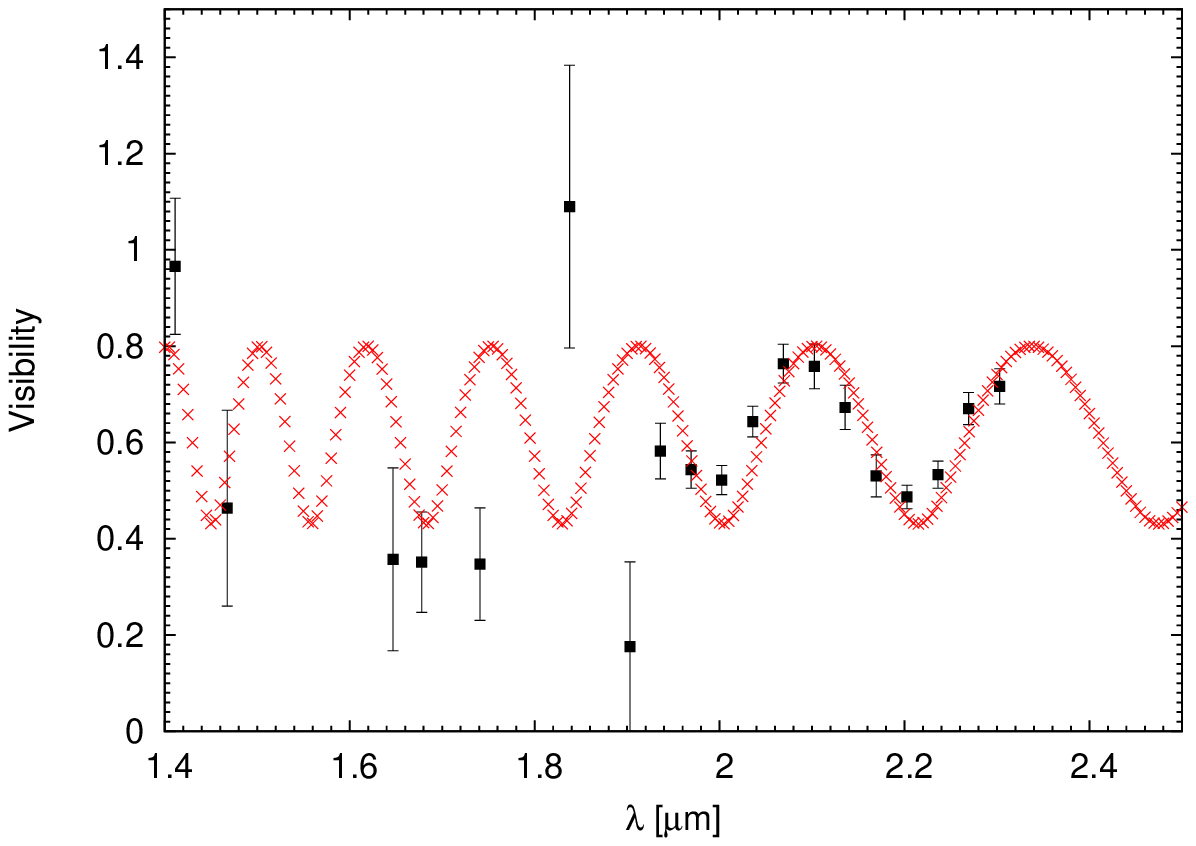}}}
\caption{\label{Vis_PAR1891_1} Visibilities of $\theta^1$ Ori C observed on 05/10/2010. \textit{Left:} Baseline K0-I1, 44\,m, PA $8{}^{\circ}$ \textit {Middle:} 
Baseline I1-A0, 110\,m, PA $-89{}^{\circ}$ \textit {Right:} Baseline A0-K0, 123\,m, PA $70{}^{\circ}$
}
\end{figure*}

\begin{figure*}
\parbox{18.5cm}{\parbox{6.0cm}{\includegraphics[width=6.0cm, angle=0]{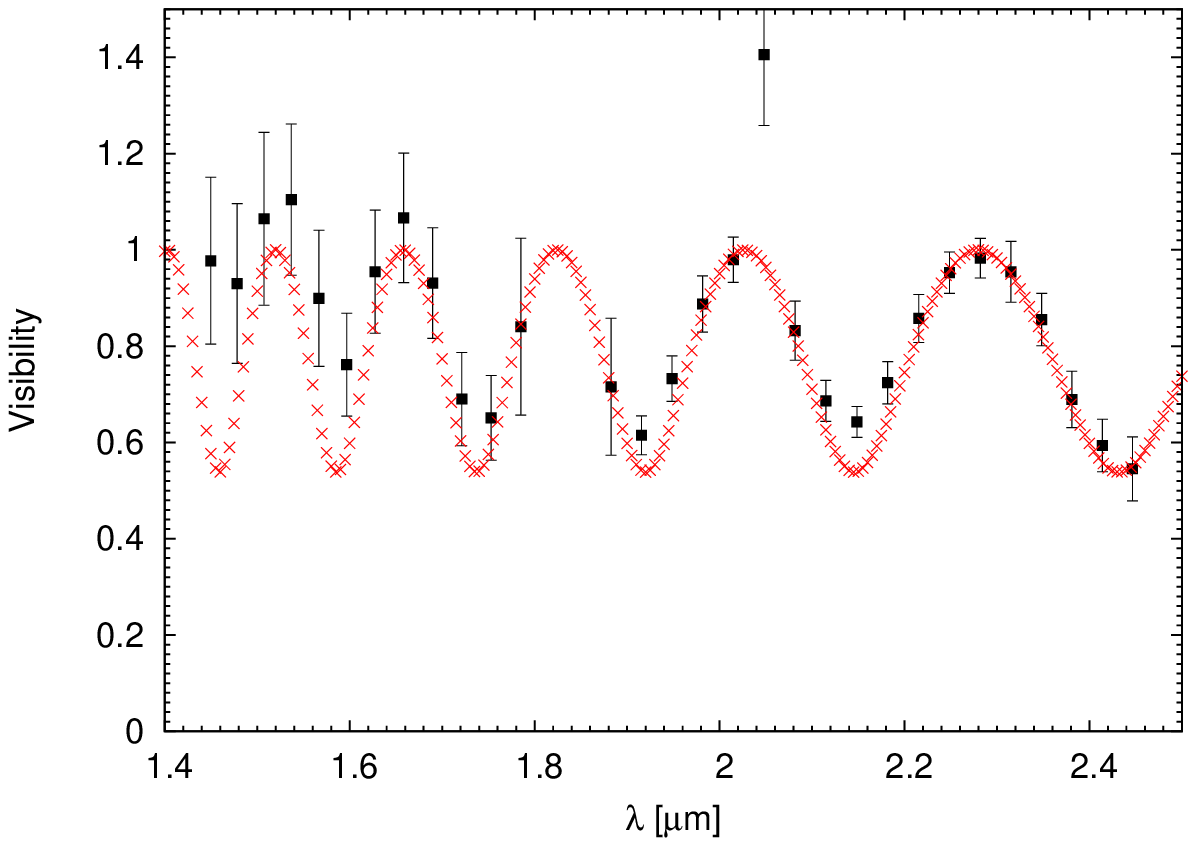}}
\parbox{6.0cm}{\includegraphics[width=6.0cm, angle=0]{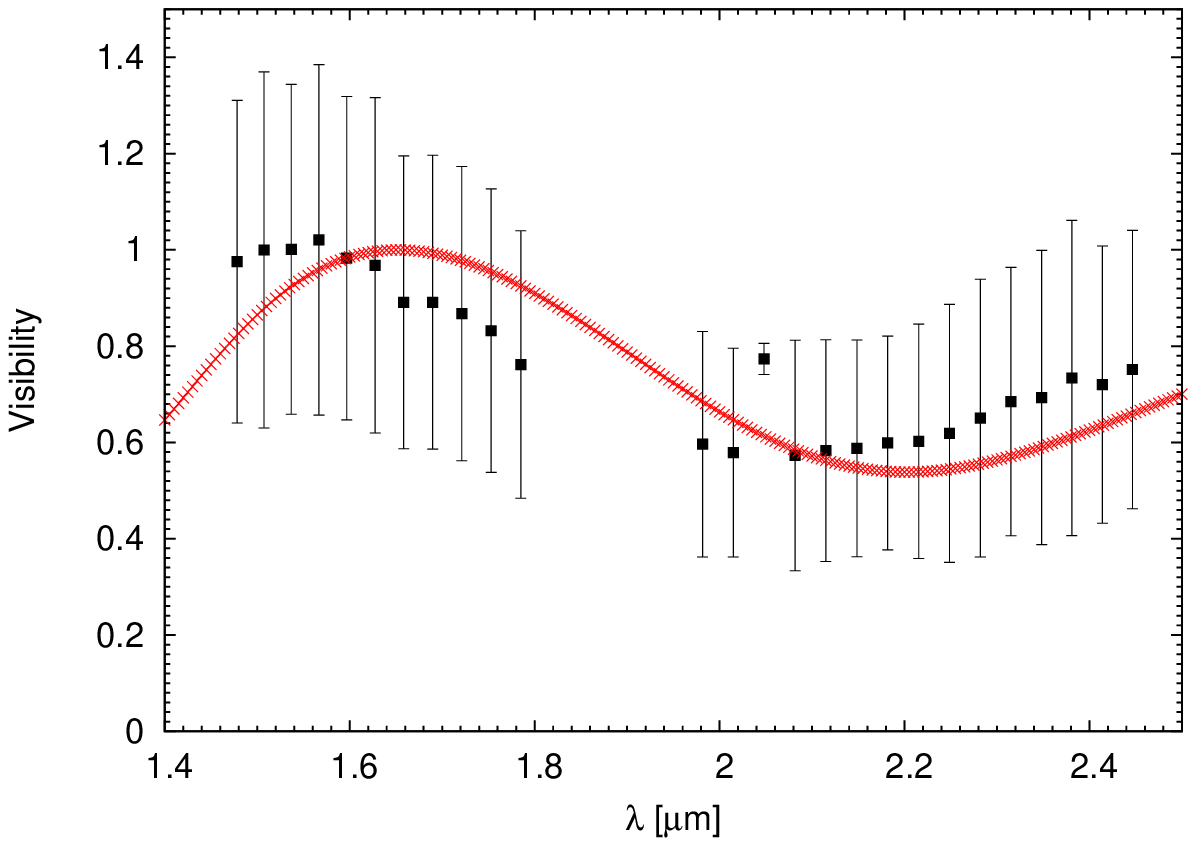}}\parbox{6.0cm}{\includegraphics[width=6.0cm, angle=0]{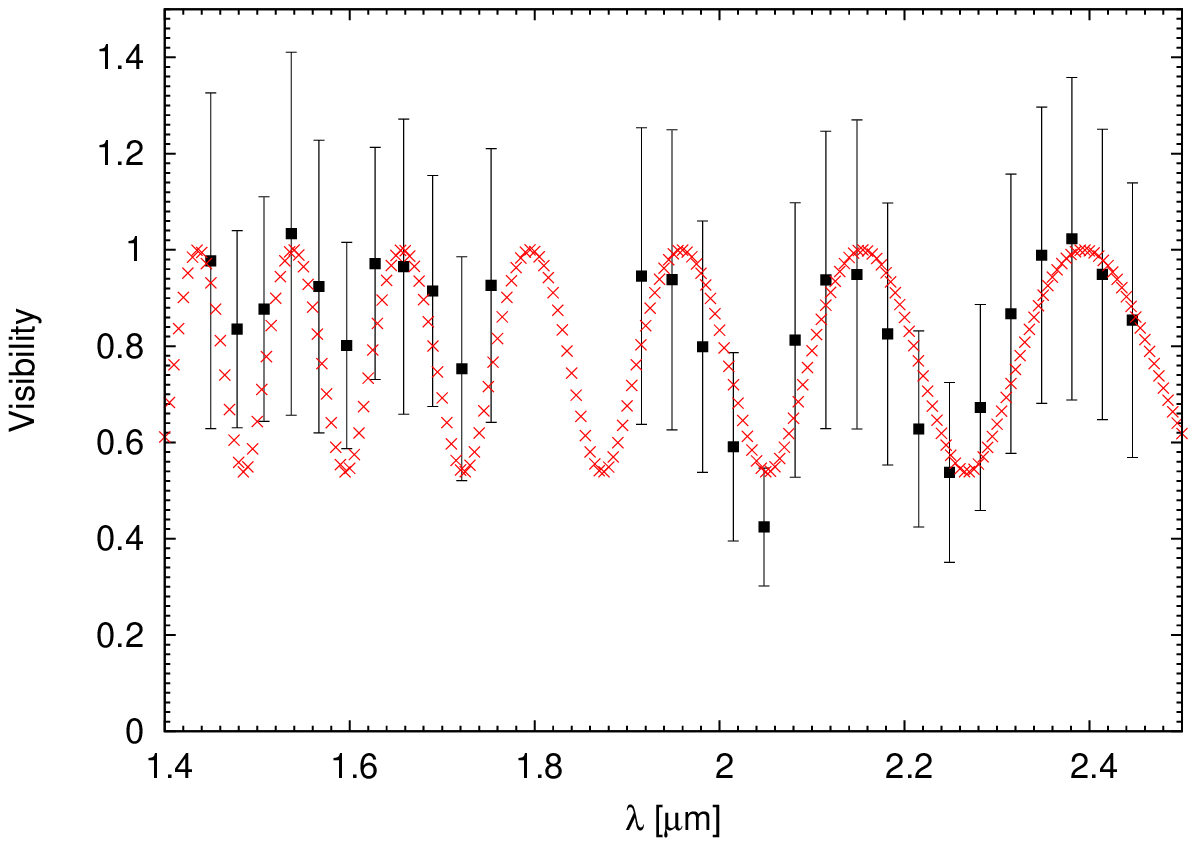}}}
\caption{\label{Vis_PAR1891_2} Visibilities of $\theta^1$ Ori C observed on 26/12/2010. \textit{Left:} Baseline K0-G1, 87\,m, PA $29{}^{\circ}$ \textit {Middle:} 
Baseline G1-A0, 88\,m, PA $-65{}^{\circ}$ \textit {Right:} Baseline A0-K0, 128\,m, PA $72{}^{\circ}$
}
\end{figure*}

\begin{figure*}
\parbox{18.5cm}{\parbox{6.0cm}{\includegraphics[width=6.0cm, angle=0]{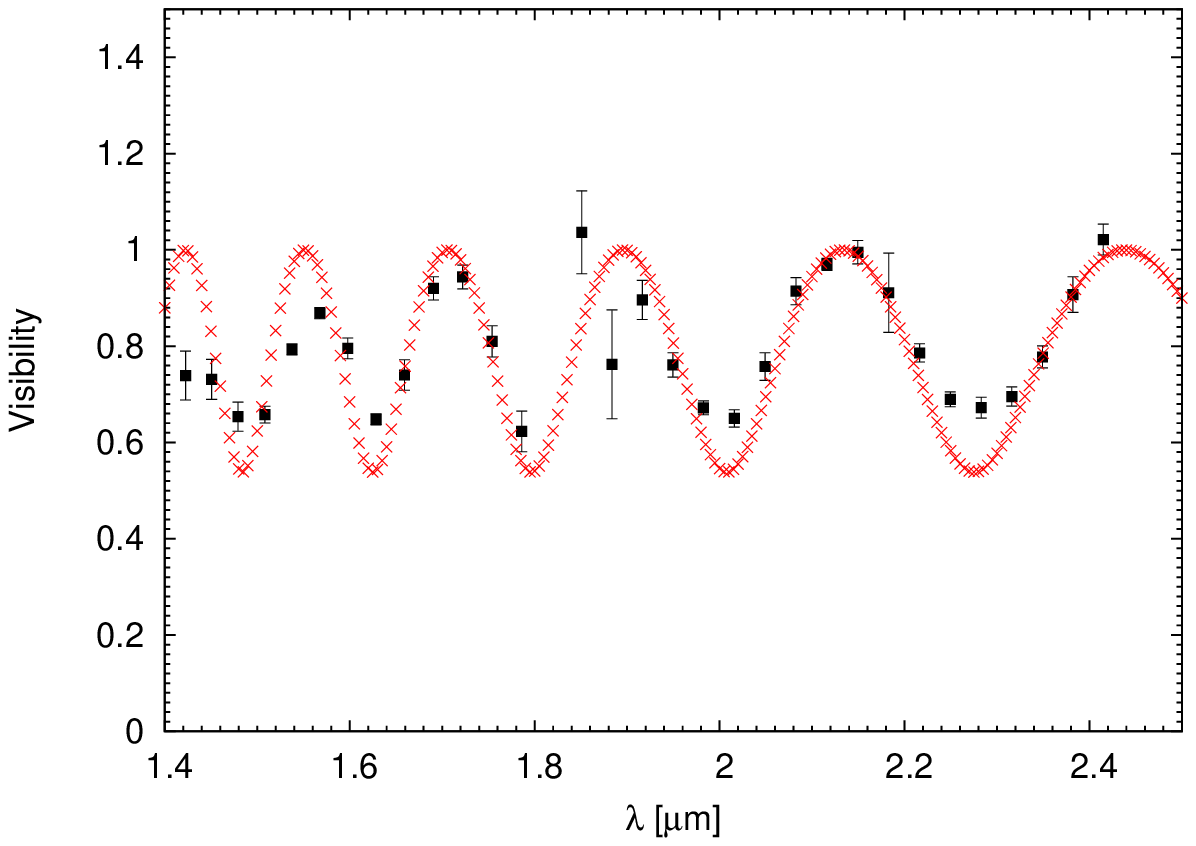}}
\parbox{6.0cm}{\includegraphics[width=6.0cm, angle=0]{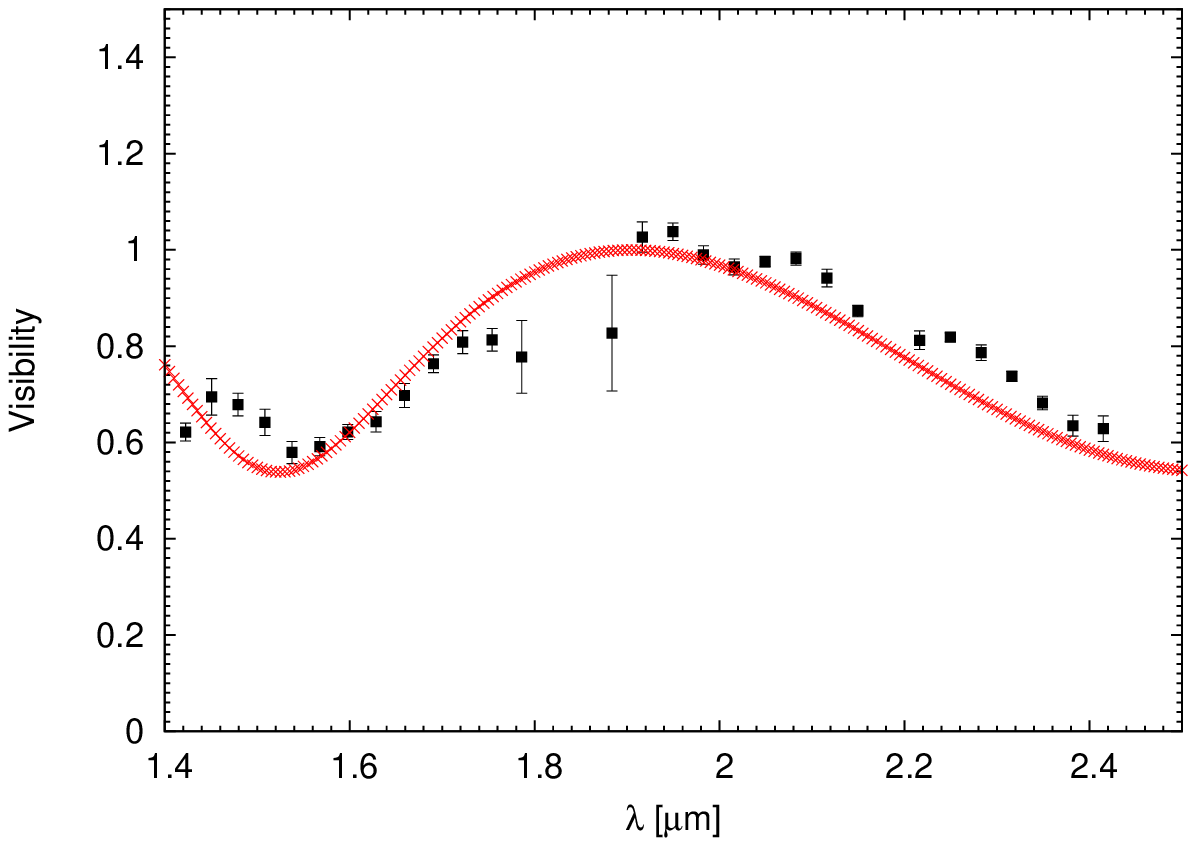}}\parbox{6.0cm}{\includegraphics[width=6.0cm, angle=0]{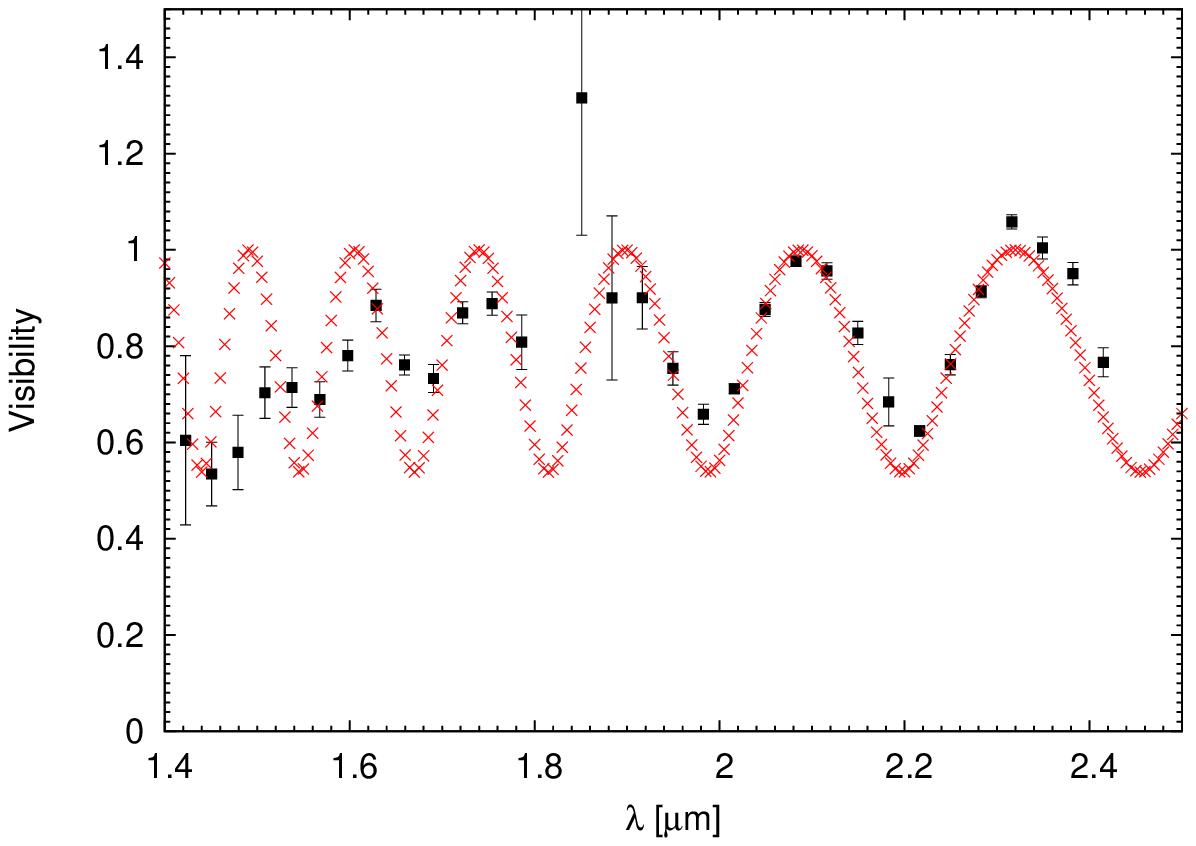}}}
\caption{\label{Vis_PAR1891_3} Visibilities of $\theta^1$ Ori C observed on 27/12/2010. \textit{Left:} Baseline K0-G1, 84\,m, PA $22{}^{\circ}$ \textit {Middle:} 
Baseline G1-A0, 90\,m, PA $-67{}^{\circ}$ \textit {Right:} Baseline A0-K0, 122\,m, PA $70{}^{\circ}$
}
\end{figure*}

\begin{figure*}
\parbox{18.5cm}{\parbox{6.0cm}{\includegraphics[width=6.0cm, angle=0]{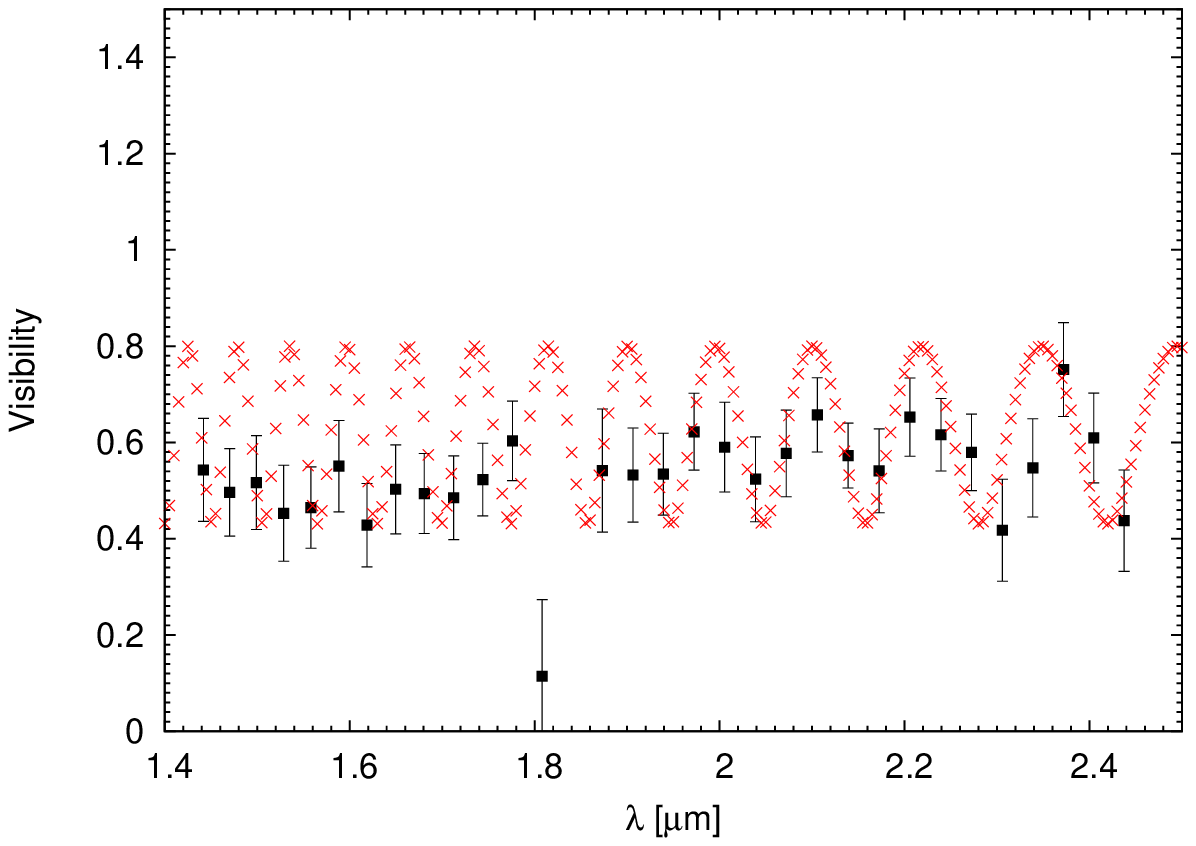}}
\parbox{6.0cm}{\includegraphics[width=6.0cm, angle=0]{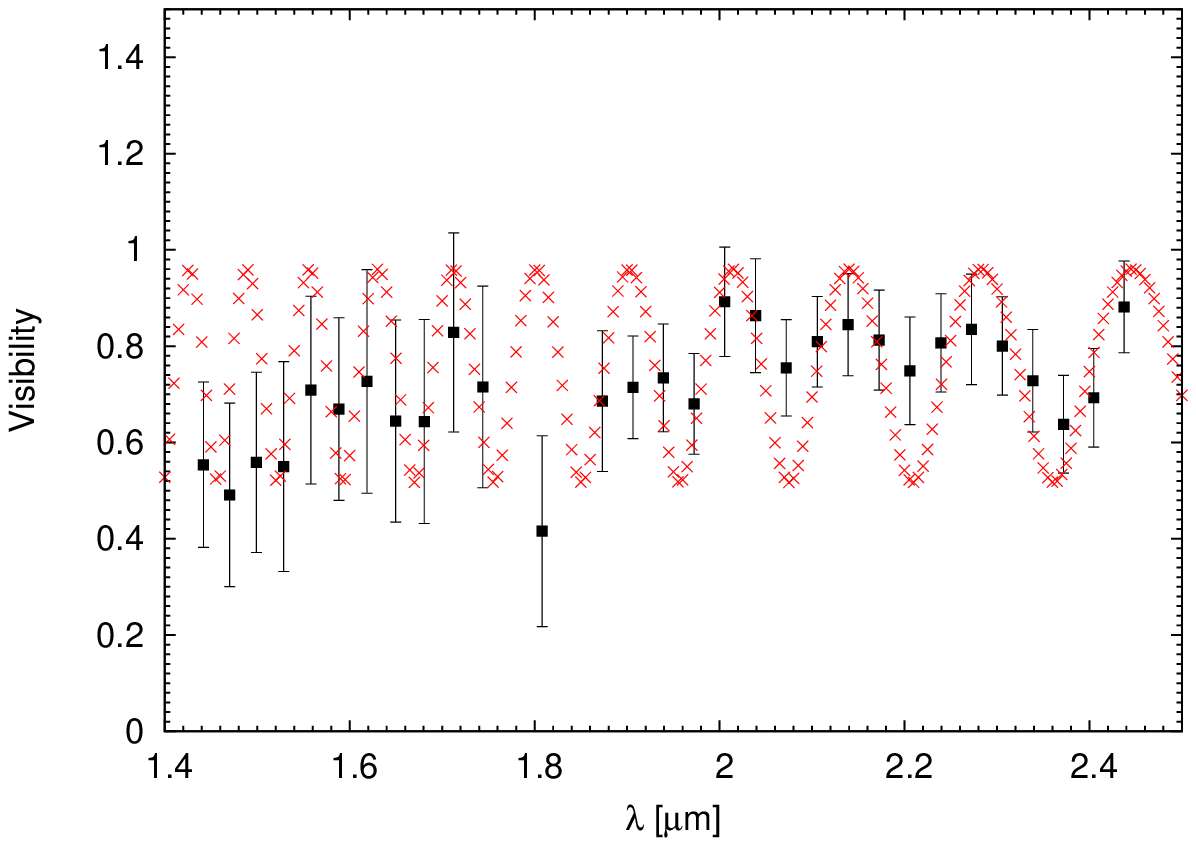}}\parbox{6.0cm}{\includegraphics[width=6.0cm, angle=0]{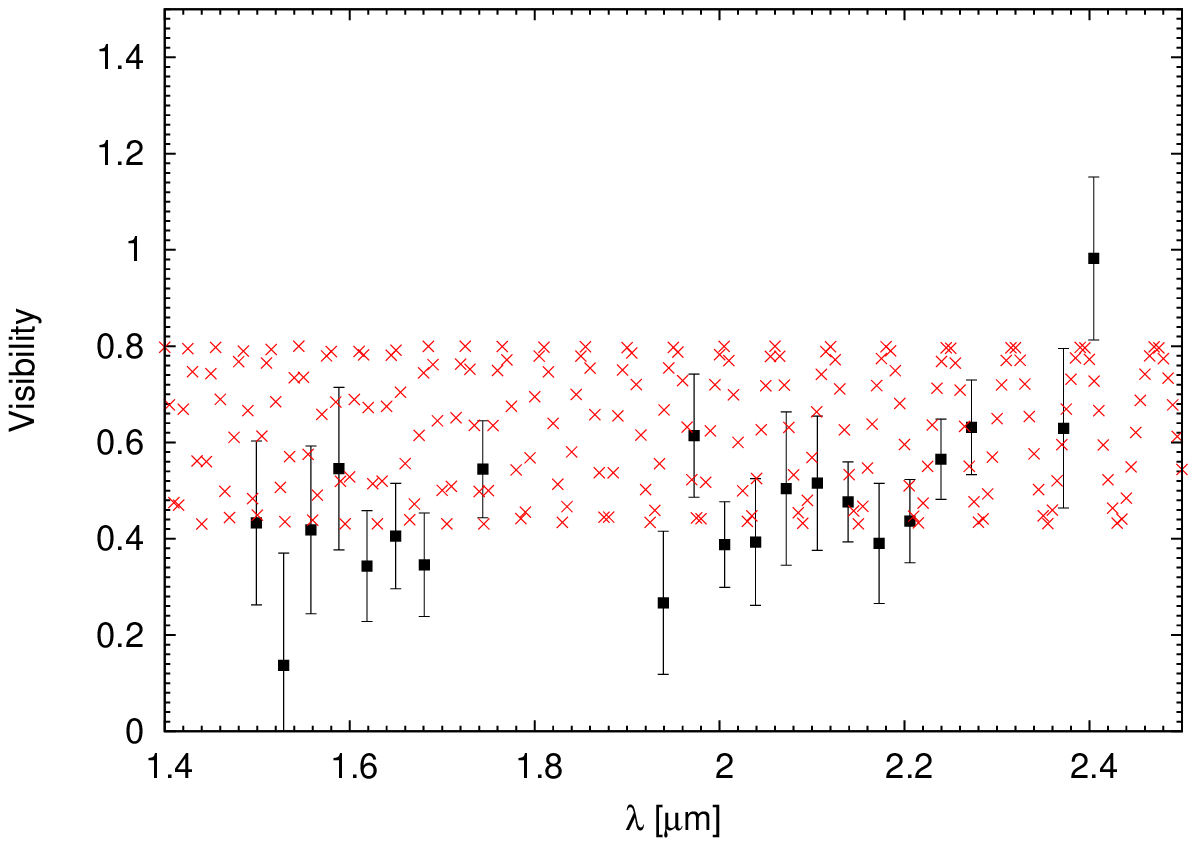}}}
\caption{\label{Vis_PAR1865_1} Visibilities of $\theta^1$ Ori A observed on 14/12/2010.  \textit{Left:} Baseline K0-I1, 43\,m, PA $180{}^{\circ}$ \textit {Middle:} 
Baseline G1-I1, 40\,m, PA $-150{}^{\circ}$ \textit {Right:} Baseline G1-K0, 81\,m, PA $-166{}^{\circ}$
}
\end{figure*}

\begin{figure*}
\parbox{18.5cm}{\parbox{6.0cm}{\includegraphics[width=6.0cm, angle=0]{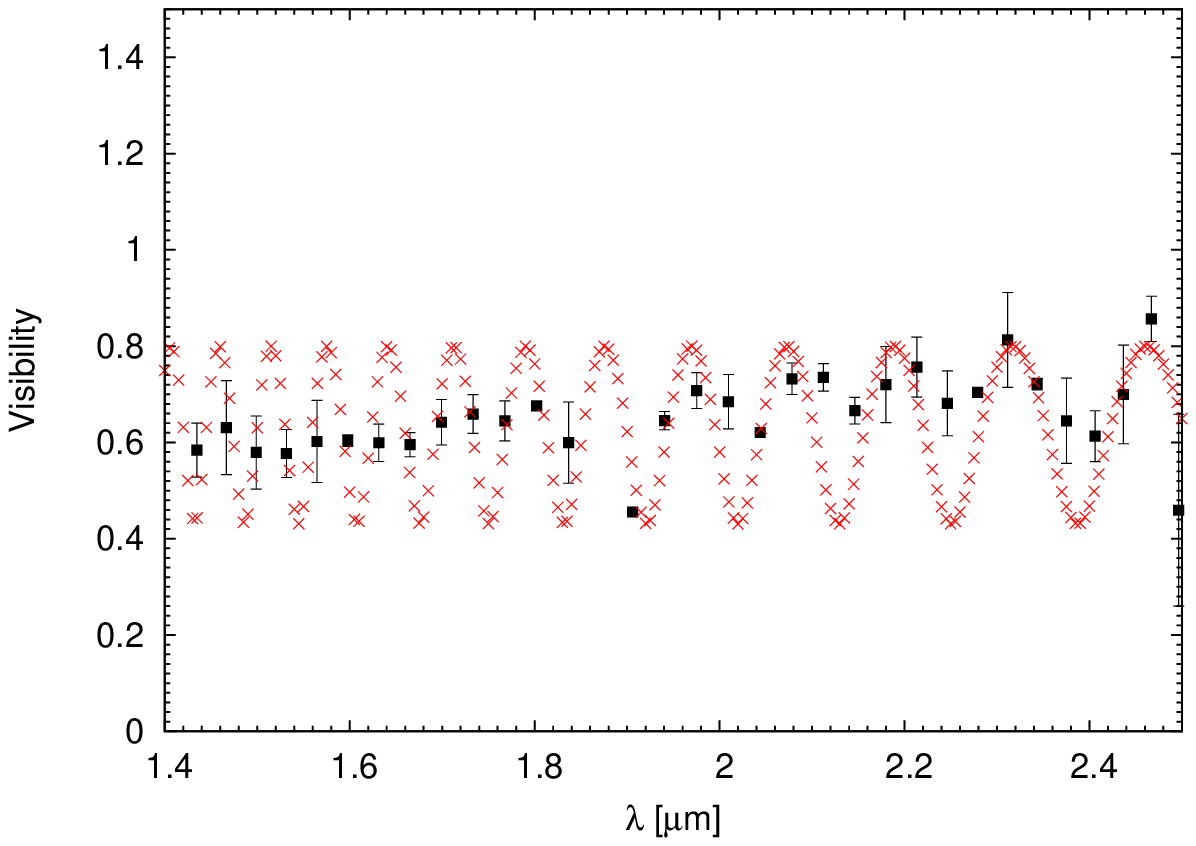}}
\parbox{6.0cm}{\includegraphics[width=6.0cm, angle=0]{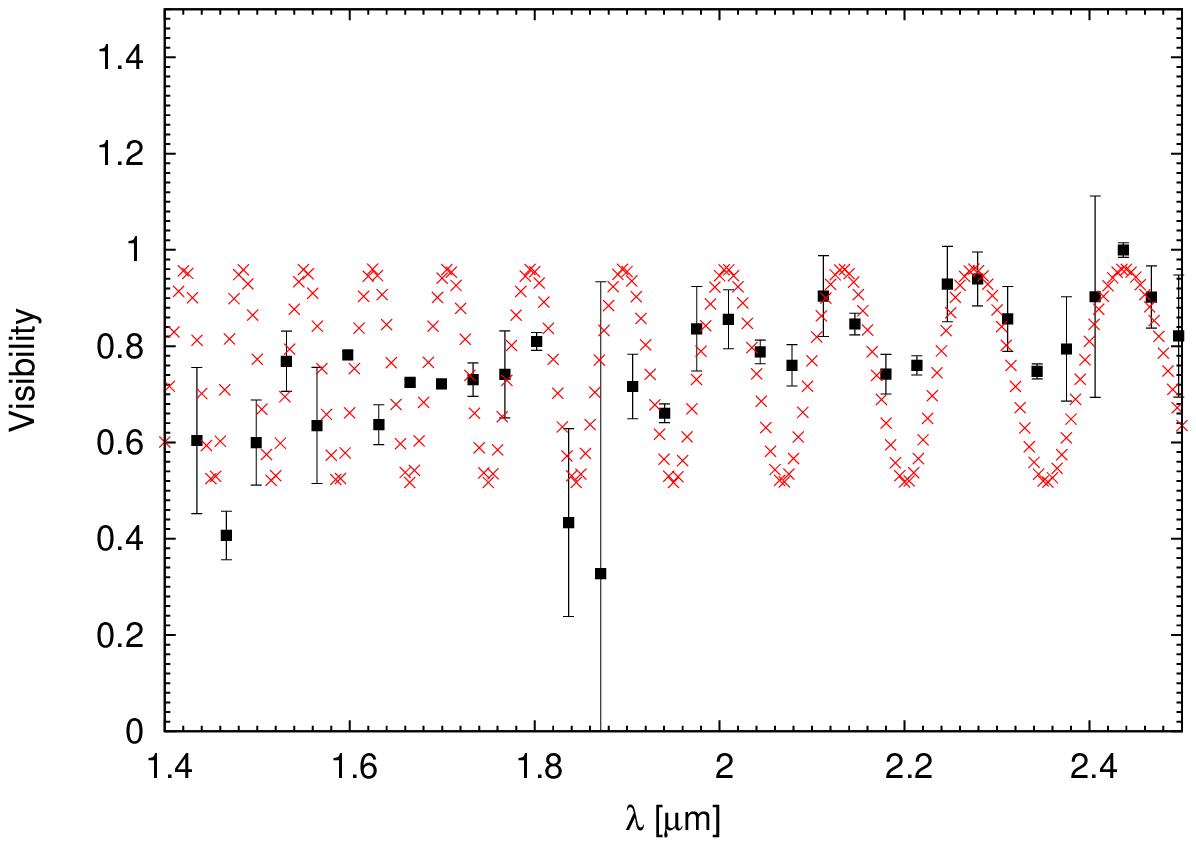}}\parbox{6.0cm}{\includegraphics[width=6.0cm, angle=0]{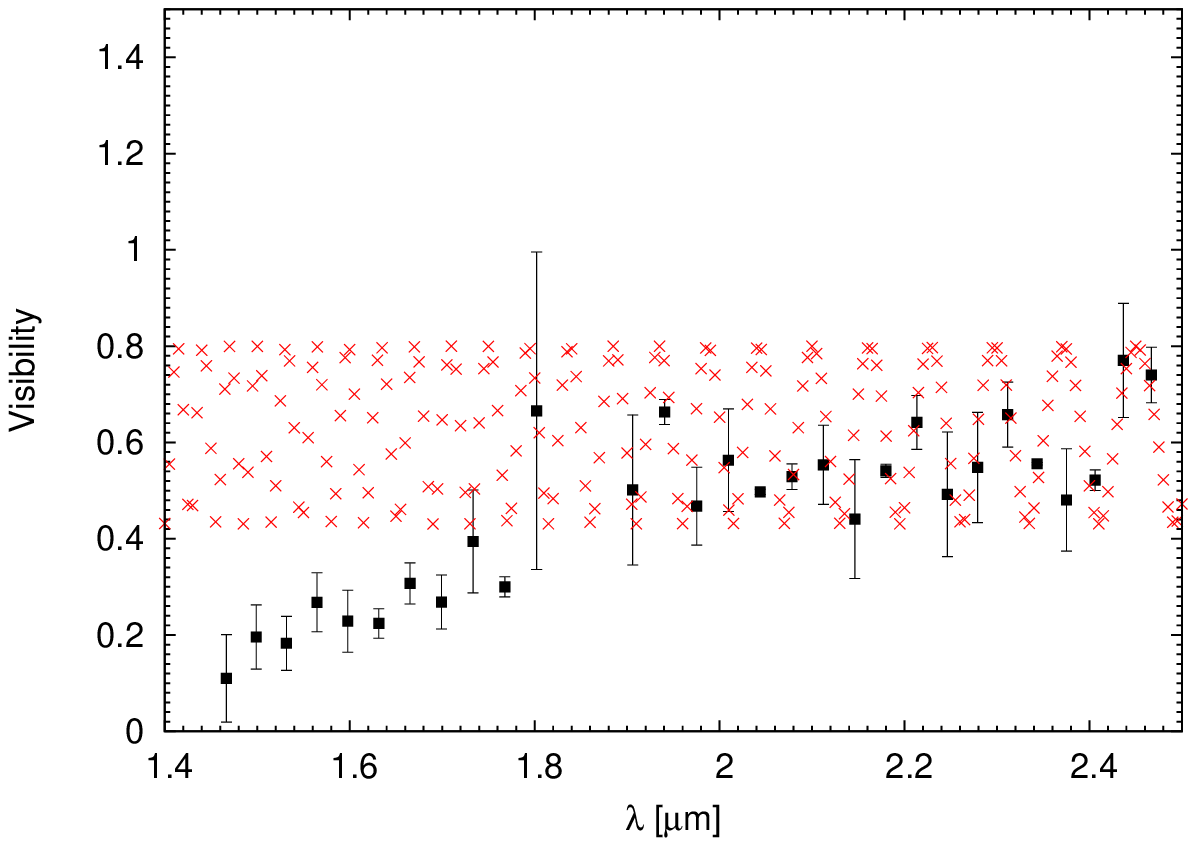}}}
\caption{\label{Vis_PAR1865_2} Visibilities of $\theta^1$ Ori A observed on 29/10/2011. \textit{Left:} Baseline K0-I1, 43\,m, PA $177{}^{\circ}$ \textit {Middle:} 
Baseline G1-I1, 41\,m, PA $-147{}^{\circ}$ \textit {Right:} Baseline G1-K0, 81\,m, PA $-163{}^{\circ}$  
}
\end{figure*}

\begin{figure*}
\parbox{18.5cm}{\parbox{6.0cm}{\includegraphics[width=6.0cm, angle=0]{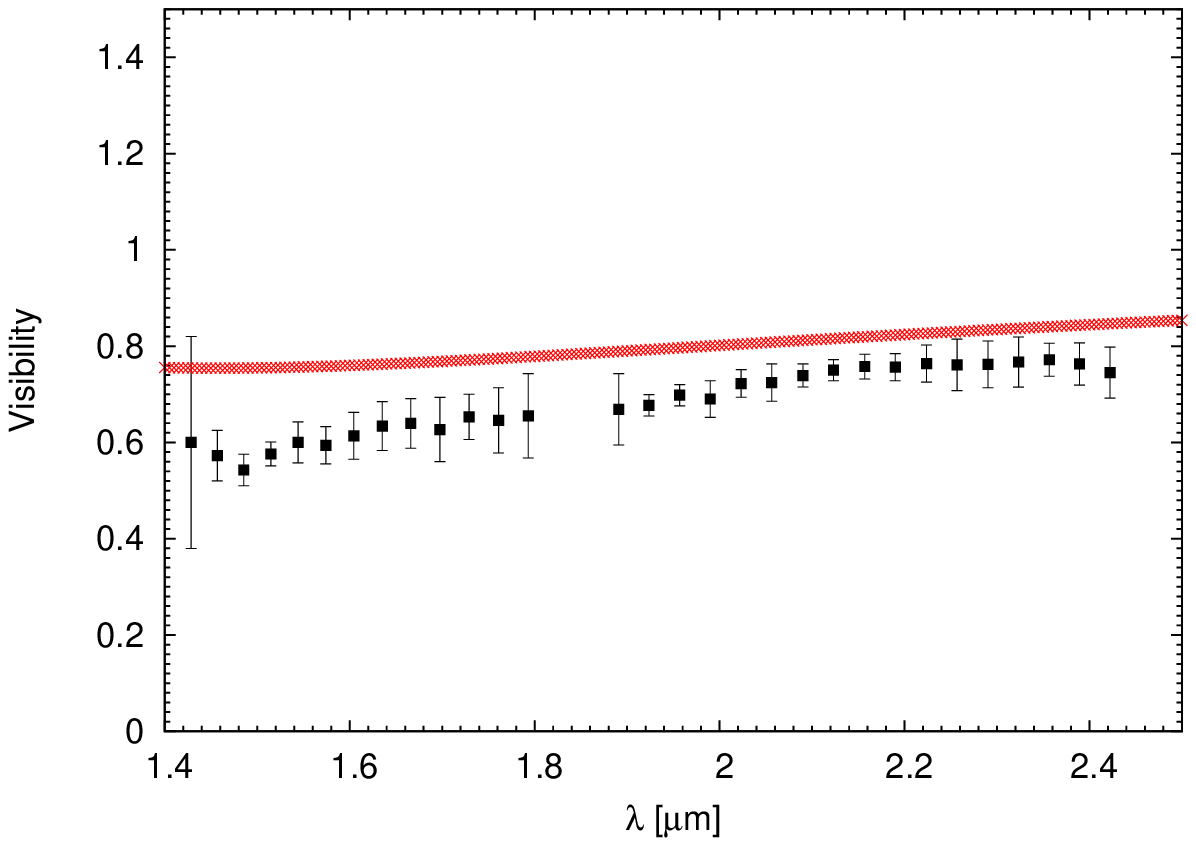}}
\parbox{6.0cm}{\includegraphics[width=6.0cm, angle=0]{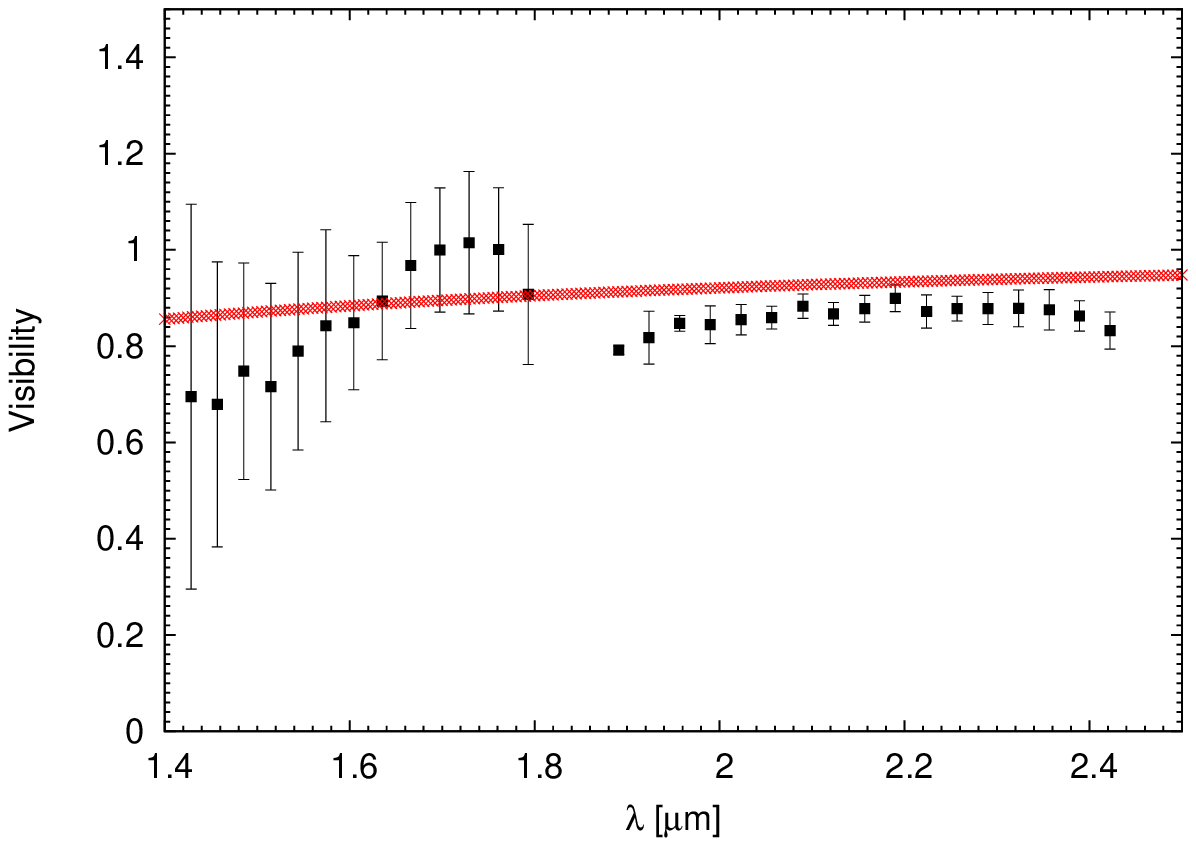}}\parbox{6.0cm}{\includegraphics[width=6.0cm, angle=0]{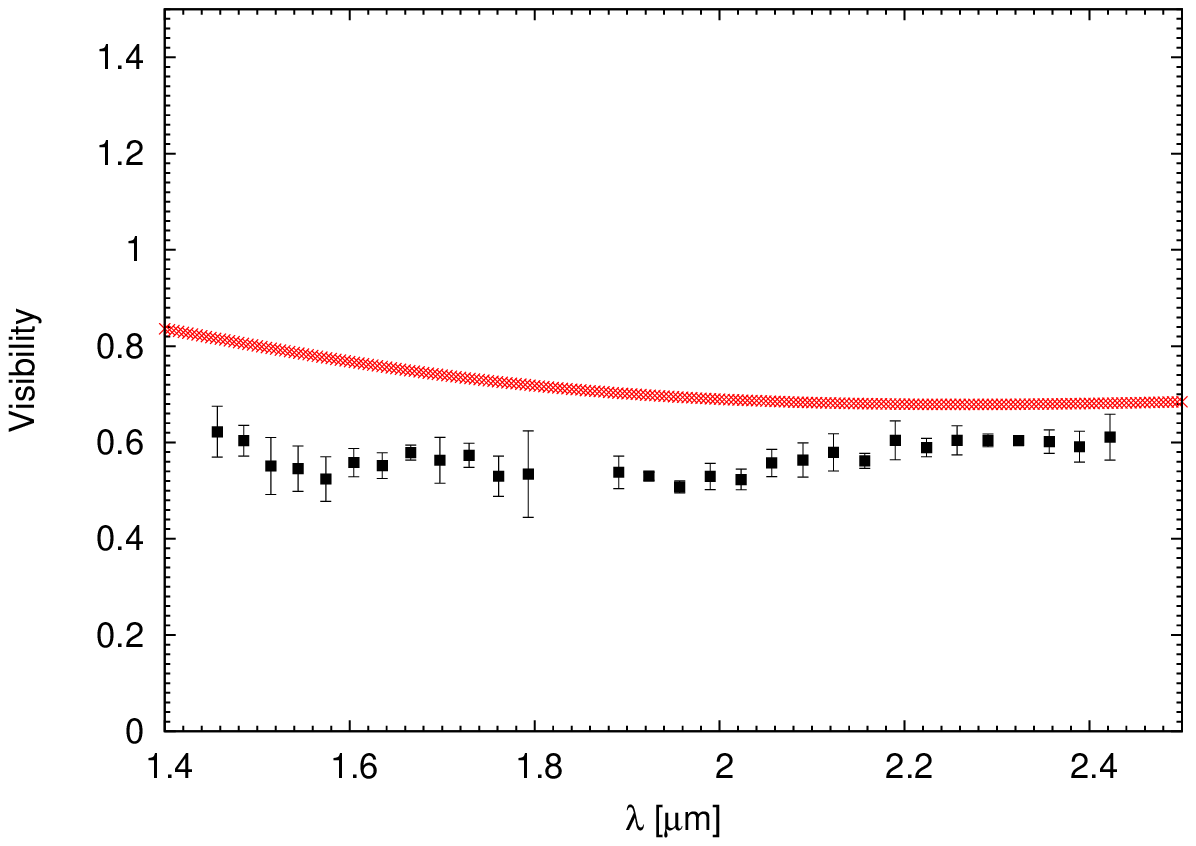}}}
\caption{\label{Vis_PAR1889_2} Visibilities of $\theta^1$~Ori~D observed on 17/01/2011.  \textit{Left:} Baseline U1-U3, 102\,m, PA $38{}^{\circ}$ \textit {Middle:} 
Baseline U1-U4, 130\,m, PA $64{}^{\circ}$ \textit {Right:} Baseline U3-U4, 58\,m, PA $-113{}^{\circ}$
}
\end{figure*}
\begin{figure*}
\parbox{18.5cm}{\parbox{6.0cm}{\includegraphics[width=6.0cm, angle=0]{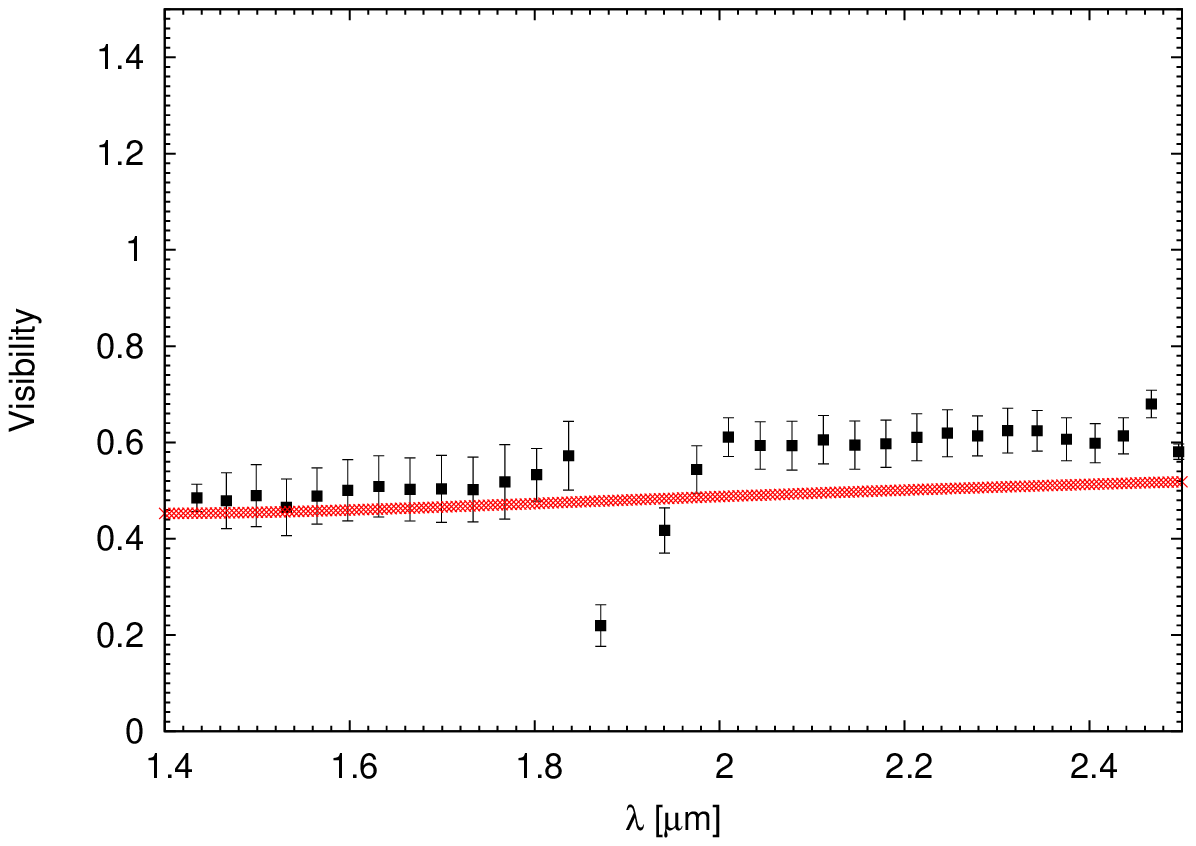}}
\parbox{6.0cm}{\includegraphics[width=6.0cm, angle=0]{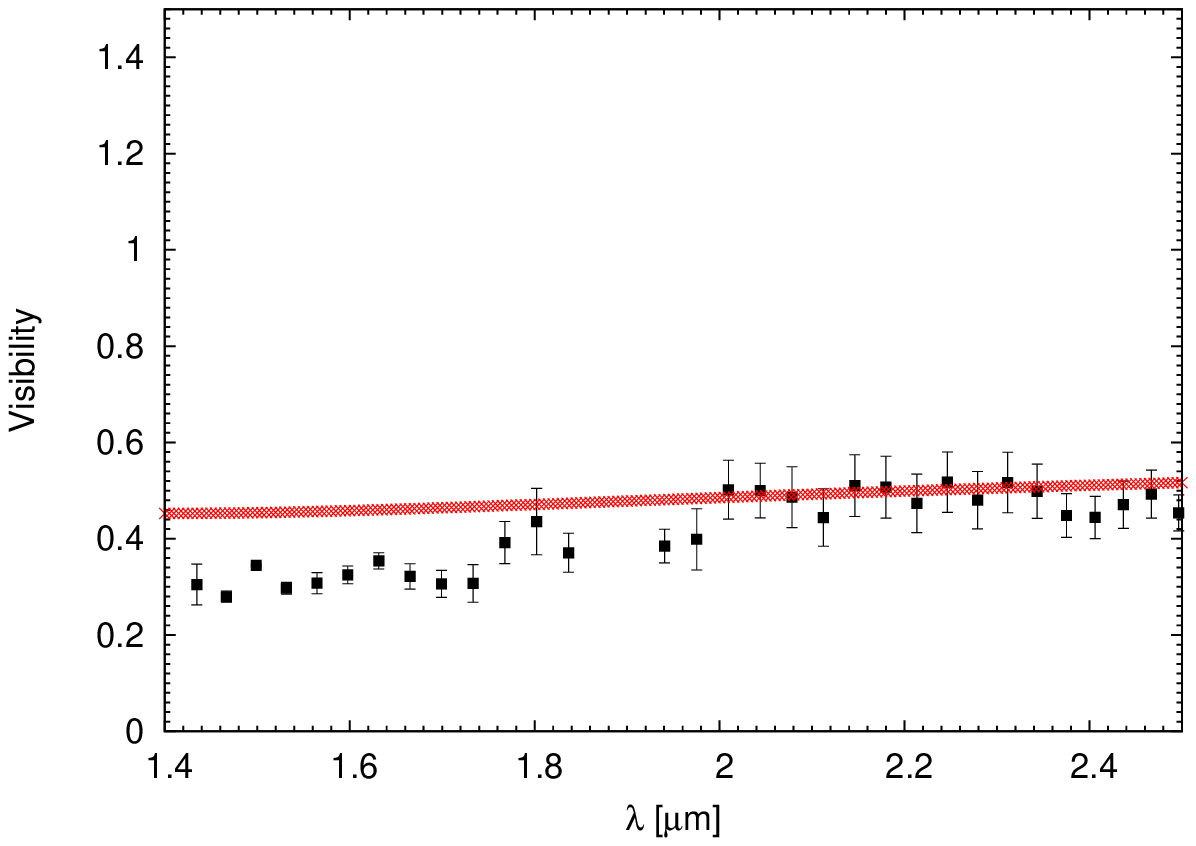}}\parbox{6.0cm}{\includegraphics[width=6.0cm, angle=0]{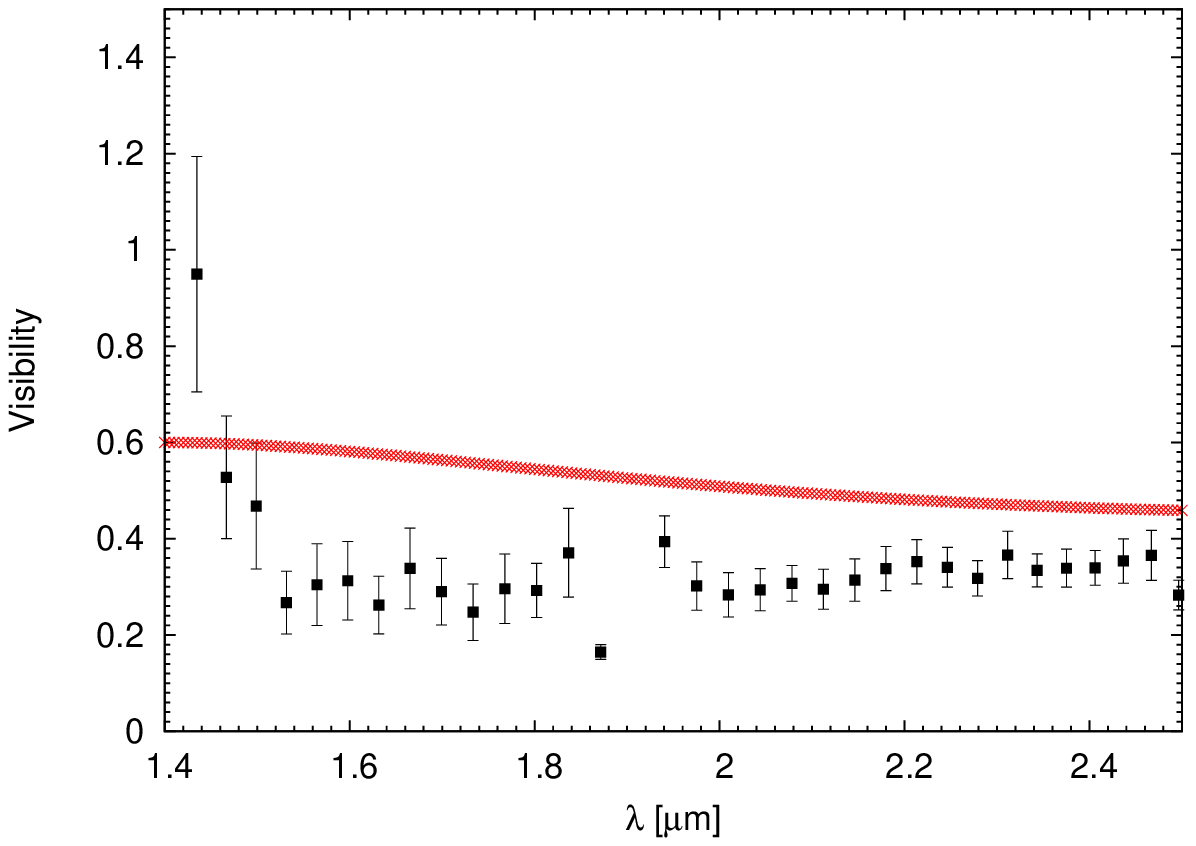}}}
\caption{\label{Vis_PAR1889_3} Visibilities of $\theta^1$~Ori~D observed on 30/10/2011. \textit{Left:} Baseline K0-I1, 43\,m, PA $175{}^{\circ}$ \textit {Middle:} 
Baseline G1-I1, 38\,m, PA $-156{}^{\circ}$ \textit {Right:} Baseline G1-K0, 79\,m, PA $-171{}^{\circ}$  
}
\end{figure*}

\begin{figure*}
\parbox{18.5cm}{\parbox{6.0cm}{\includegraphics[width=6.0cm, angle=0]{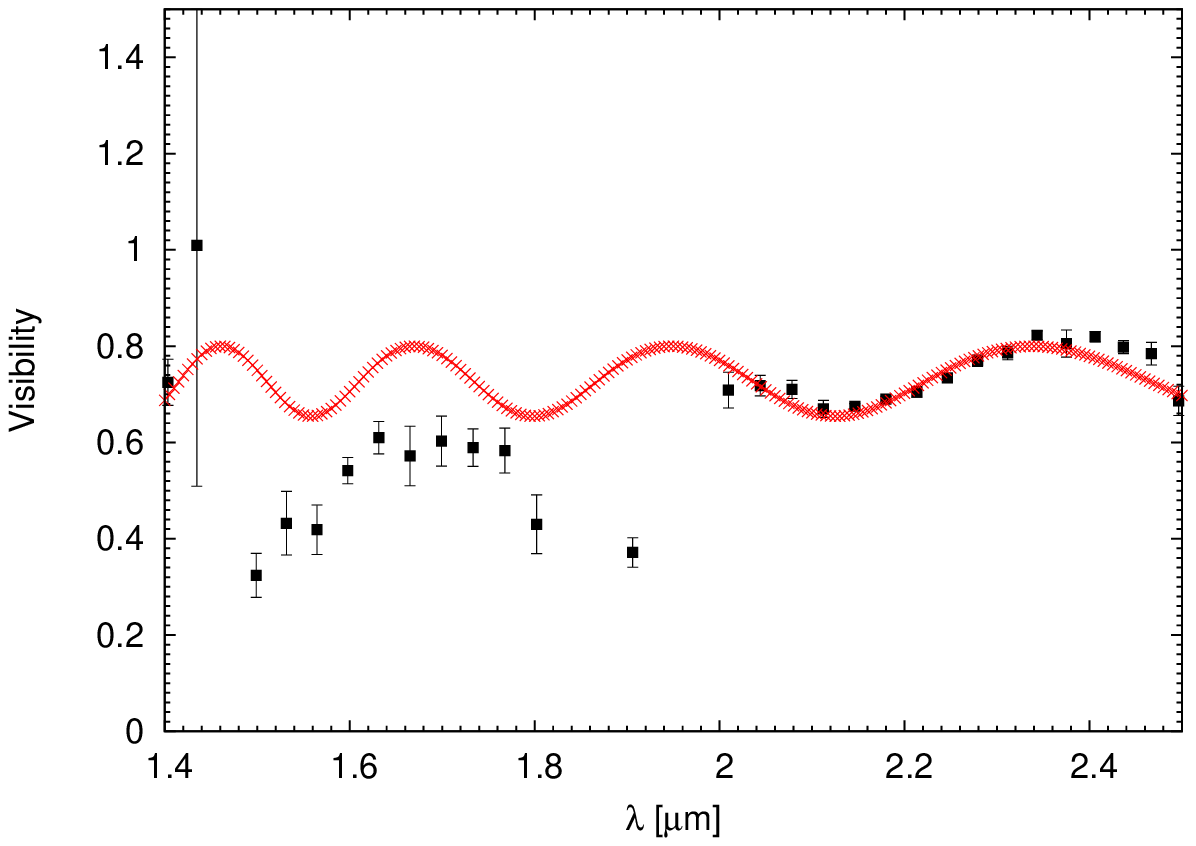}}
\parbox{6.0cm}{\includegraphics[width=6.0cm, angle=0]{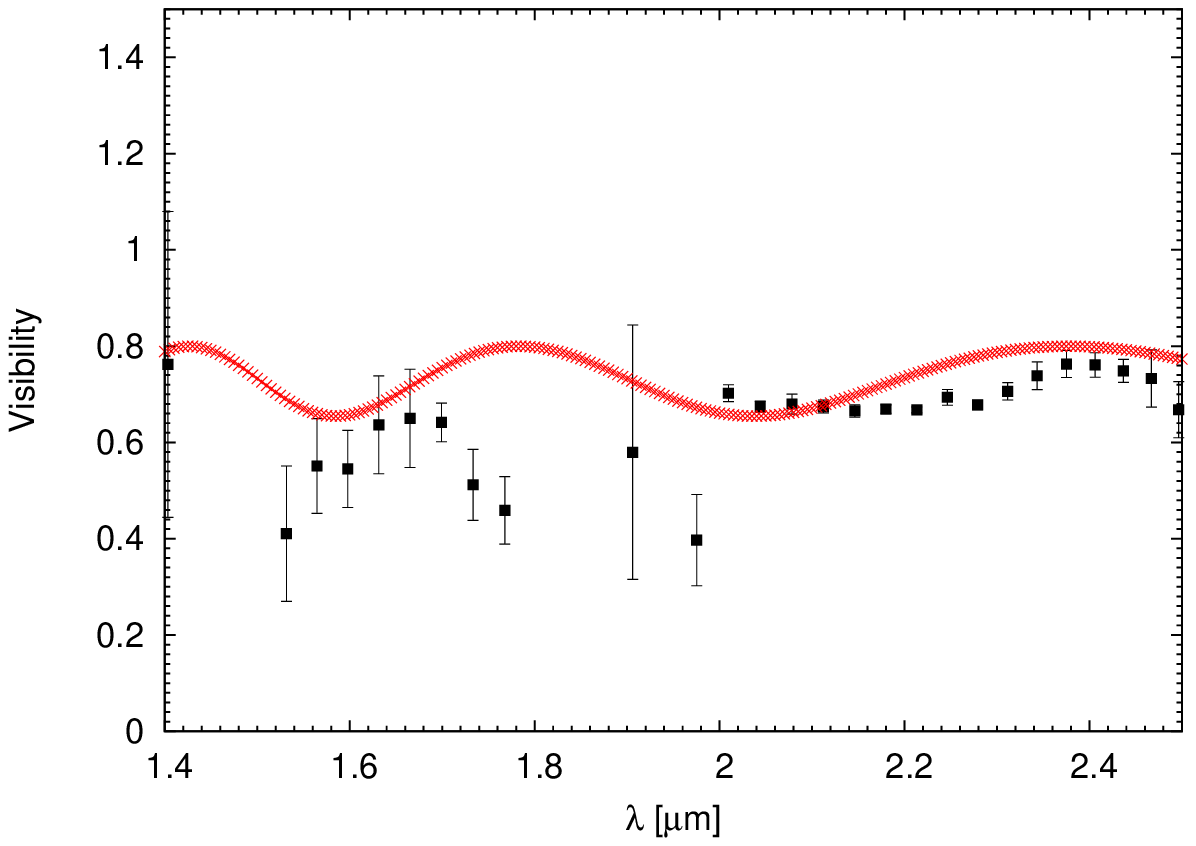}}\parbox{6.0cm}{\includegraphics[width=6.0cm, angle=0]{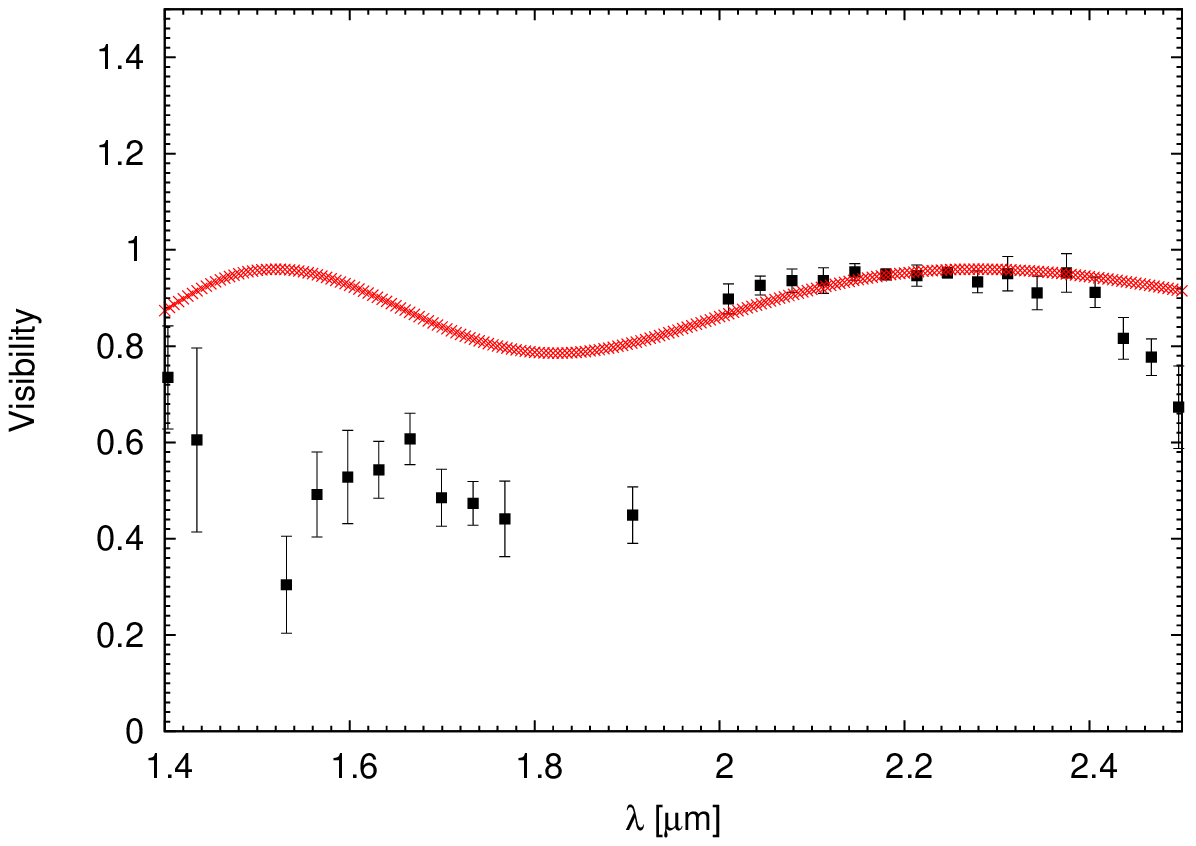}}}
\caption{\label{Vis_PAR2074_1} Visibilities of NU Ori observed on 31/12/2011.\textit{Left:} Baseline K0-A1, 127\,m, PA $-115{}^{\circ}$ \textit {Middle:} 
Baseline G1-A1, 80\,m, PA $107{}^{\circ}$ \textit {Right:} Baseline G1-K0, 86\,m, PA $-153{}^{\circ}$  
}
\end{figure*}

\begin{figure*}
\parbox{18.5cm}{\parbox{6.0cm}{\includegraphics[width=6.0cm, angle=0]{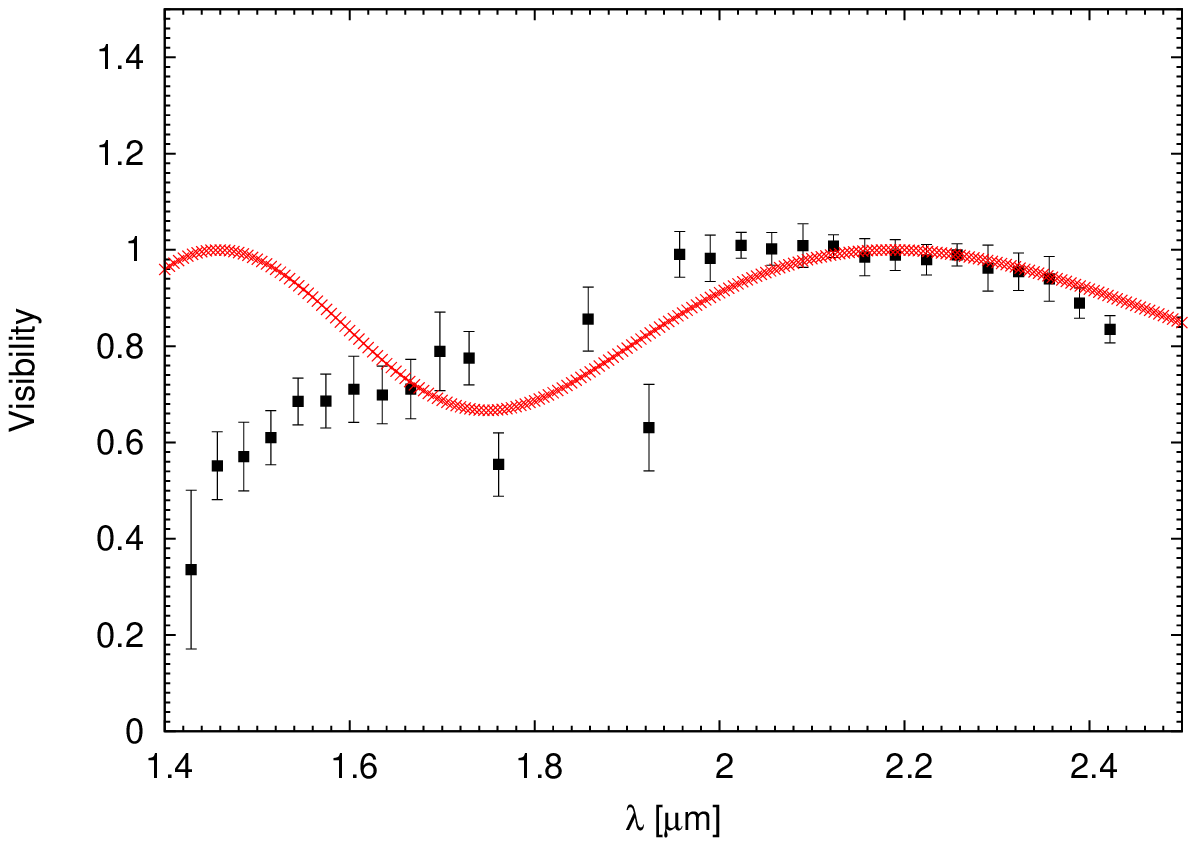}}
\parbox{6.0cm}{\includegraphics[width=6.0cm, angle=0]{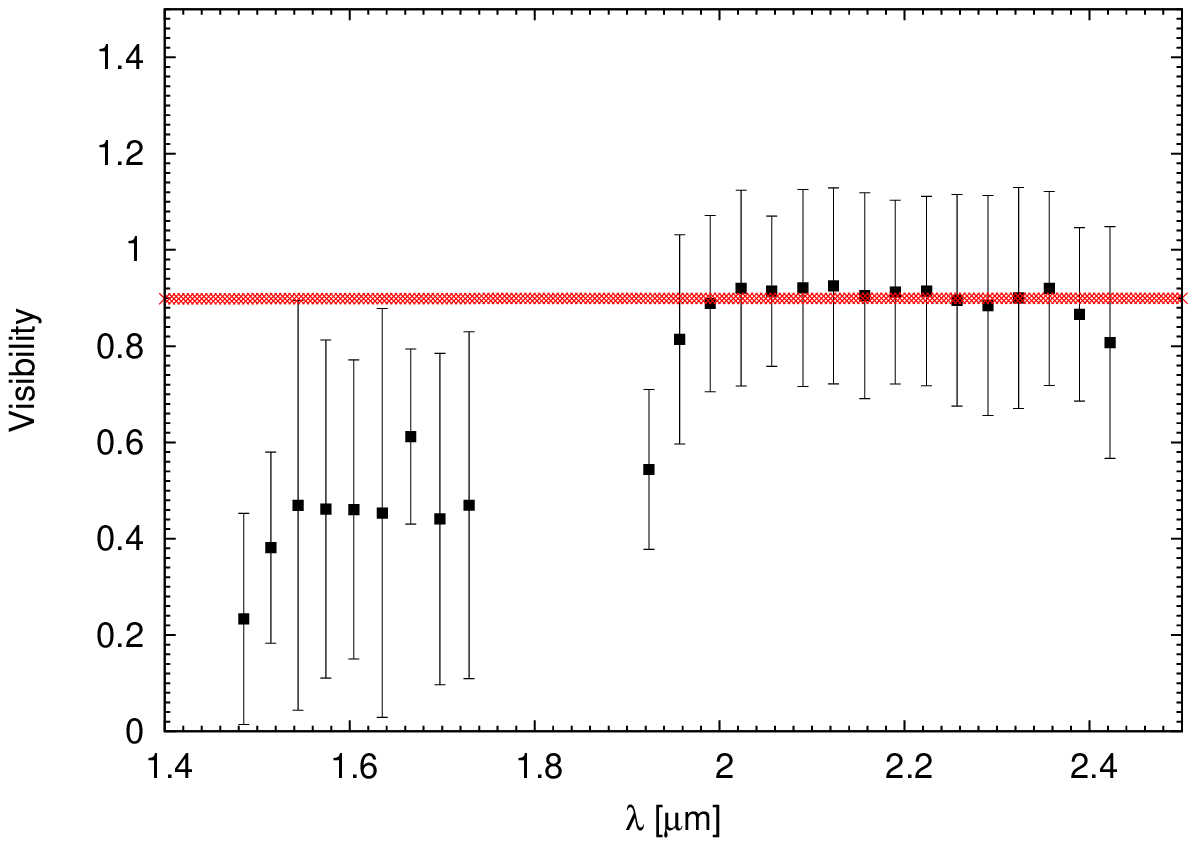}}\parbox{6.0cm}{\includegraphics[width=6.0cm, angle=0]{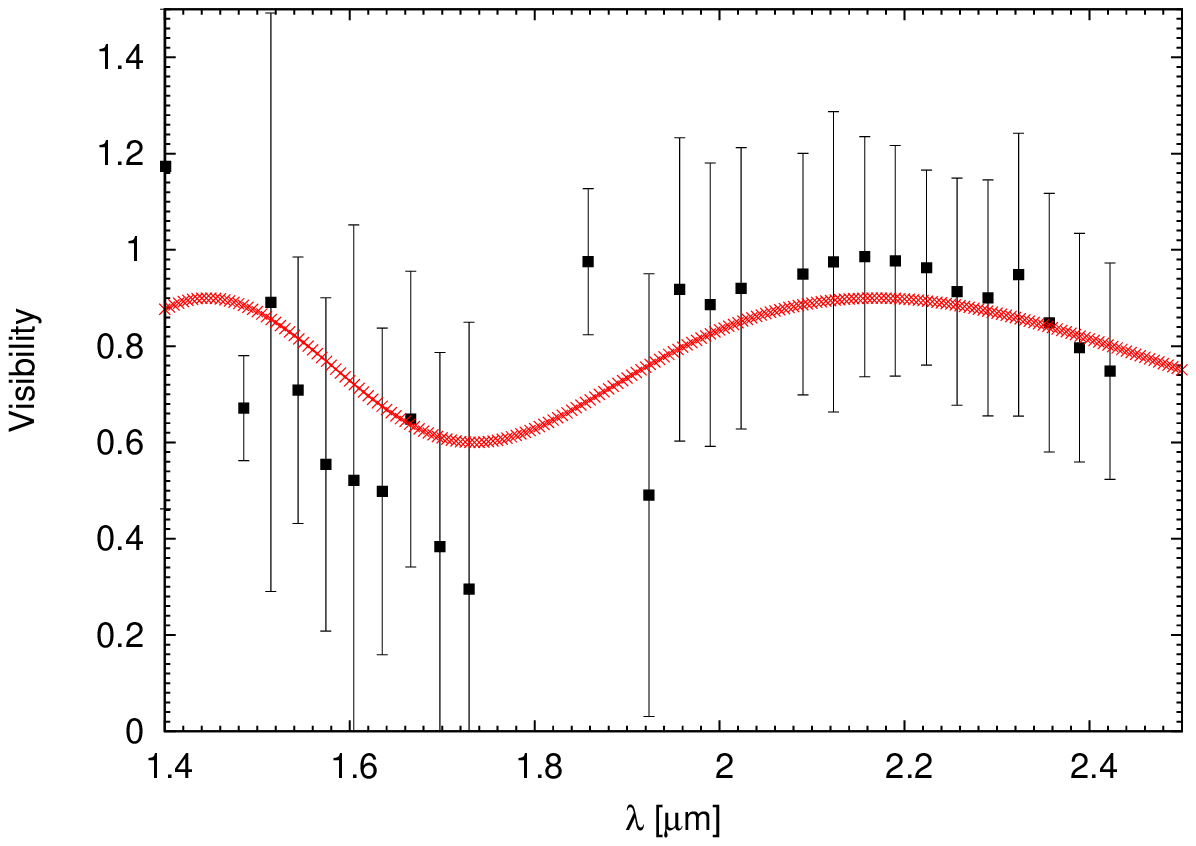}}}
\caption{\label{Vis_PAR2074_2} Visibilities of NU Ori observed on 26/03/2011. \textit{Left:} Baseline K0-G1, 90\,m, PA $-144{}^{\circ}$ \textit {Middle:} 
Baseline G1-A0, 68\,m, PA $-52{}^{\circ}$ \textit {Right:} Baseline A0-K0, 111\,m, PA $-106{}^{\circ}$ }
\end{figure*}  

\begin{figure*}
\parbox{18.5cm}{\parbox{6.0cm}{\includegraphics[width=6.0cm, angle=0]{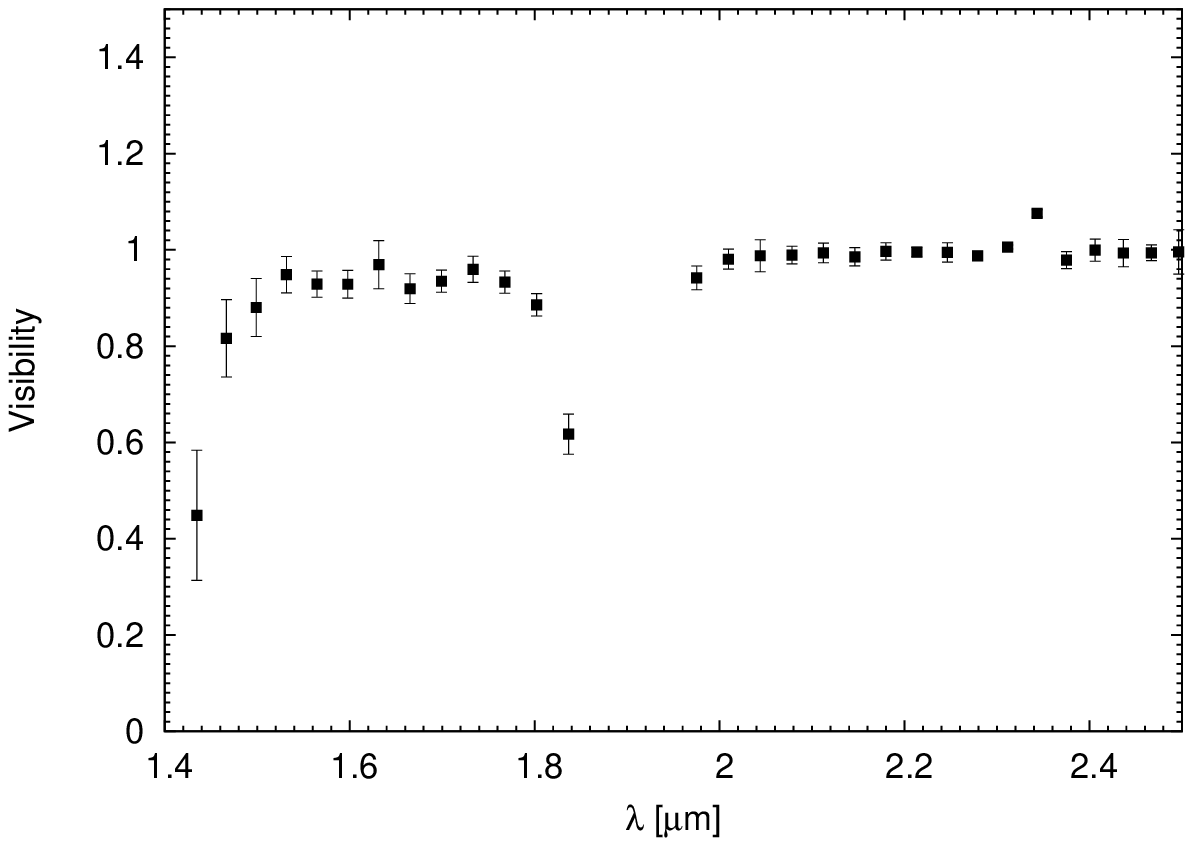}}
\parbox{6.0cm}{\includegraphics[width=6.0cm, angle=0]{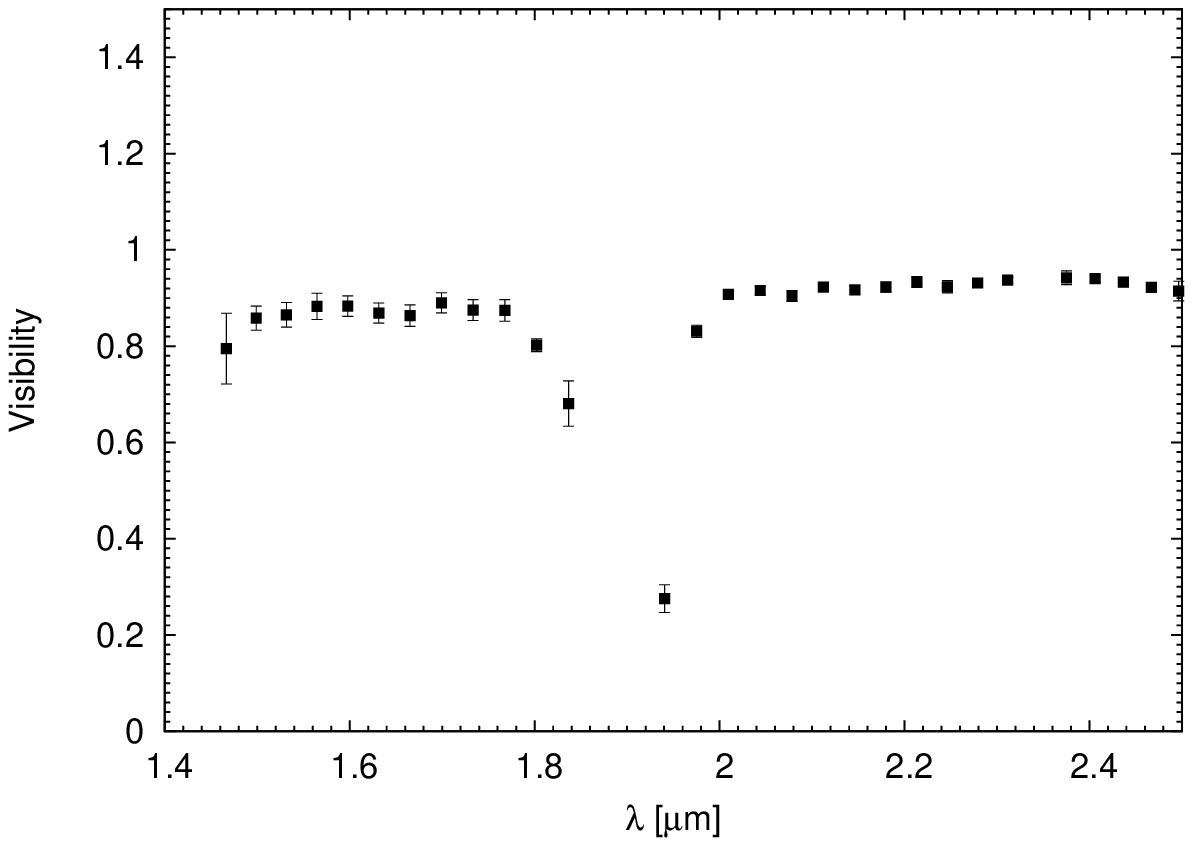}}\parbox{6.0cm}{\includegraphics[width=6.0cm, angle=0]{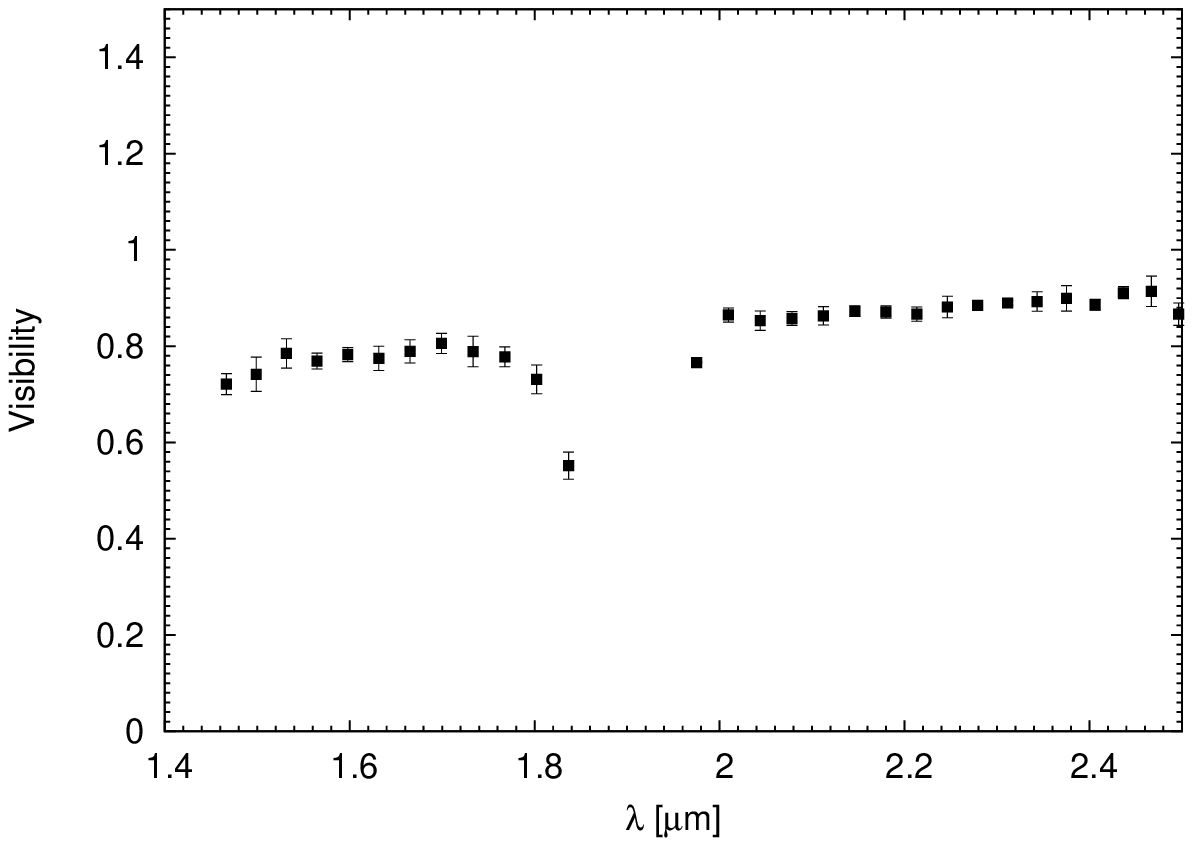}}}
\caption{\label{Vis_PAR1993_1} Visibilities $\theta^2$~Ori~A of observed on 31/12/2011. \textit{Left:} Baseline K0-A1, 127\,m, PA $-115{}^{\circ}$ \textit {Middle:} 
Baseline A1-G1, 80\,m, PA $107{}^{\circ}$ \textit {Right:} Baseline G1-K0, 86\,m, PA $-153{}^{\circ}$
}
\end{figure*}

\begin{figure*}
\parbox{18.5cm}{\parbox{6.0cm}{\includegraphics[width=6.0cm, angle=0]{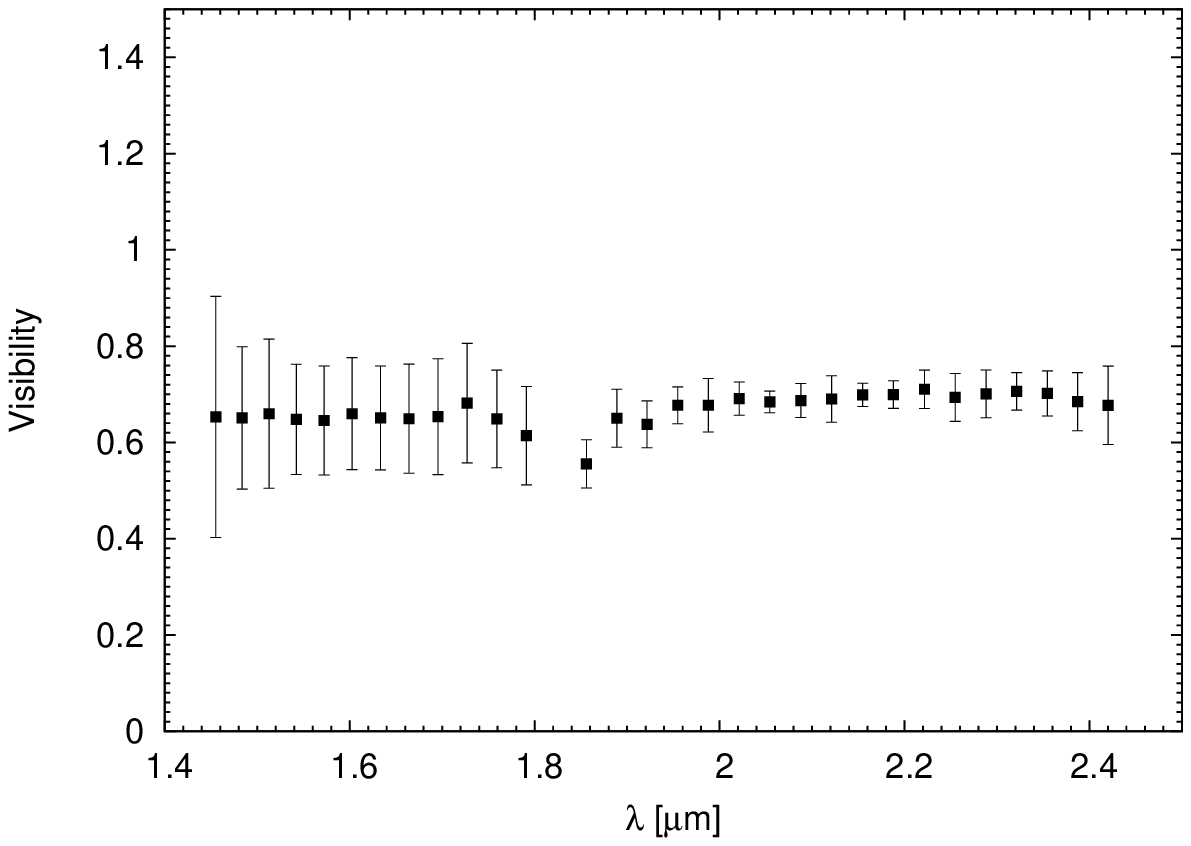}}
\parbox{6.0cm}{\includegraphics[width=6.0cm, angle=0]{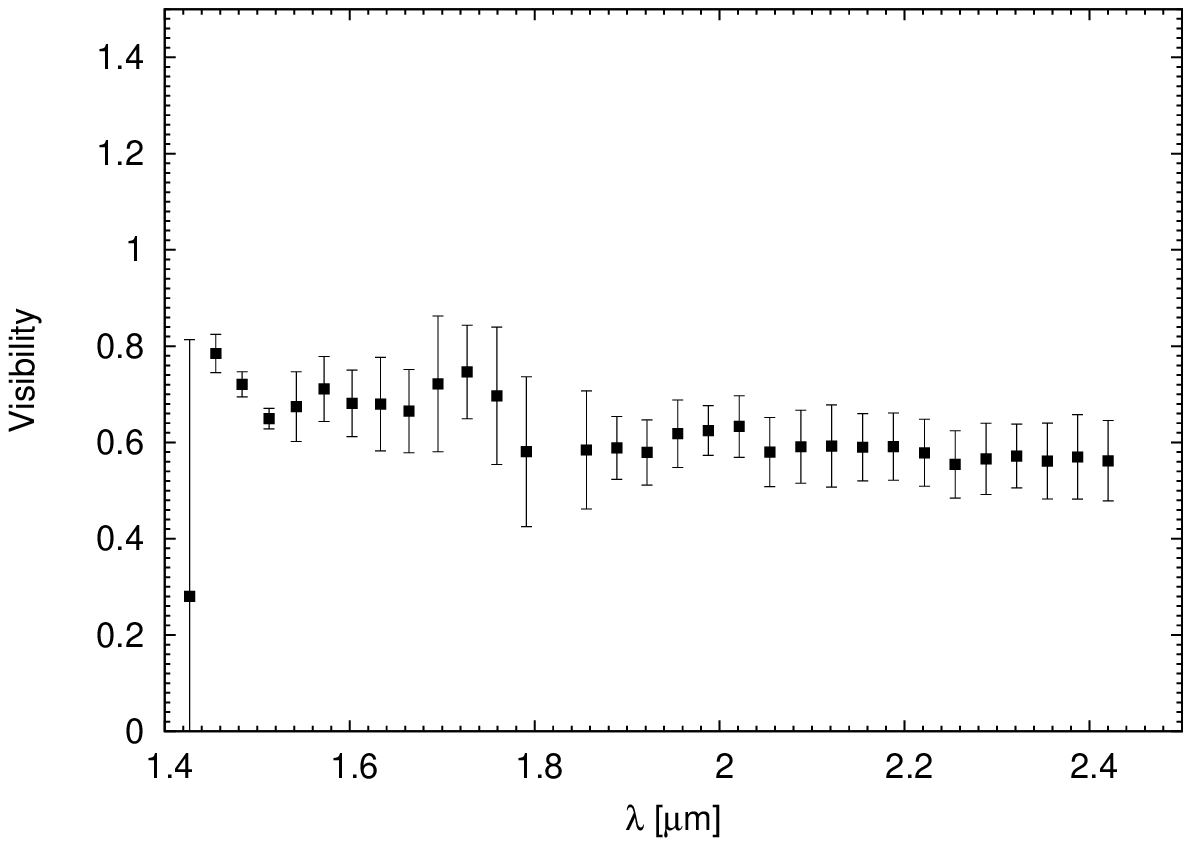}}\parbox{6.0cm}{\includegraphics[width=6.0cm, angle=0]{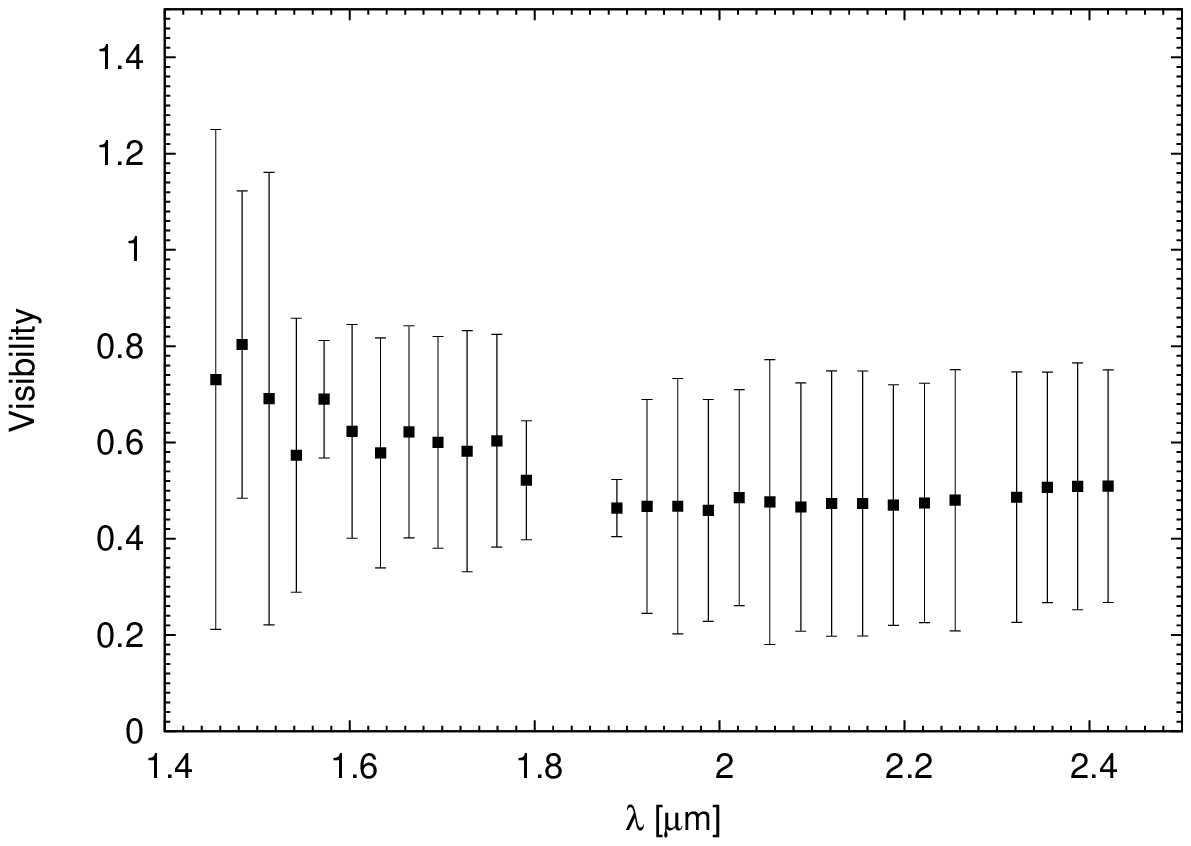}}}
\caption{\label{Vis_PAR2247_1} Visibilities of T~Ori observed on 18/01/2011. It should be mentioned again that the visibilities have not been 
calibrated absolutely. Thus, although the visibility is lower than 1 it is not sure if and how much the target is really resolved.
 \textit{Left:} Baseline U1-U3, 102\,m, PA $40{}^{\circ}$ \textit {Middle:} 
Baseline U3-U4, 53\,m, PA $116{}^{\circ}$ \textit {Right:} Baseline U4-U1, 126\,m, PA $64{}^{\circ}$}
\end{figure*}

\end{document}